\documentclass[twocolumn,twocolappendix]{aastex63}
\usepackage{multirow,booktabs}
\usepackage{upgreek}
\usepackage{amsmath}
\usepackage{xcolor}

\shorttitle{The temperature of Oph163131's disk}
\shortauthors{Flores et al.}
\graphicspath{{./}{figures/}}

\begin{document}

\title{The Anatomy of an Unusual Edge-on Protoplanetary Disk. II. \\Gas temperature and a warm outer region}

\email{caflores@hawaii.edu}
\correspondingauthor{Christian Flores}

\author[0000-0002-8591-472X]{Flores C.}
\affiliation{Institute for Astronomy, University of Hawaii at Manoa, 640 N. Aohoku Place, Hilo, HI 96720, USA}

\author[0000-0002-5092-6464]{Duch\^ ene G.}
\affiliation{Astronomy Department, University of California, Berkeley, CA 94720, USA}
\affiliation{Universit\'e Grenoble Alpes, CNRS, Institut de Plan\'etologie et d'Astrophysique, IPAG, 38000 Grenoble, France}

\author{Wolff S.}
\affiliation{University of Arizona, Tucson, AZ}

\author{Villenave M.}
\affiliation{Universit\'e Grenoble Alpes, CNRS, Institut de Plan\'etologie et d'Astrophysique, IPAG, 38000 Grenoble, France}

\author{Stapelfeldt K.}
\affiliation{Jet Propulsion Laboratory, California Institute of Technology, Mail Stop 321-100, 4800 Oak Grove Drive Pasadena, CA 91109  USA}

\author{Williams J. P.}
\affiliation{Institute for Astronomy, University of Hawaii at Manoa, 2680 Woodlawn Drive, Honolulu, HI 96822, USA}

\author{Pinte C.}
\affiliation{School of Physics and Astronomy, Monash University, Clayton Vic 3800, Australia}
\affiliation{Universit\'e Grenoble Alpes, CNRS, Institut de Plan\'etologie et d'Astrophysique, IPAG, 38000 Grenoble, France}

\author{Padgett D.}
\affiliation{Jet Propulsion Laboratory, California Institute of Technology, Mail Stop 321-100, 4800 Oak Grove Drive Pasadena, CA 91109  USA}

\author[0000-0002-8293-1428]{Connelley M. S.}
\affiliation{Institute for Astronomy, University of Hawaii at Manoa, 640 N. Aohoku Place, Hilo, HI 96720, USA}

\author{van der Plas G.}
\affiliation{Universit\'e Grenoble Alpes, CNRS, Institut de Plan\'etologie et d'Astrophysique, IPAG, 38000 Grenoble, France}

\author[0000-0002-1637-7393]{M\'enard F.}
\affiliation{Universit\'e Grenoble Alpes, CNRS, Institut de Plan\'etologie et d'Astrophysique, IPAG, 38000 Grenoble, France}

\author{Perrin M. D. }
\affiliation{Space Telescope Science Institute, Baltimore, MD 21218, USA}




\begin{abstract}

We present high-resolution $^{12}$CO and $^{13}$CO 2--1 ALMA observations, as well as optical and near-infrared spectroscopy, of the highly-inclined protoplanetary disk around SSTC2D\,J163131.2-242627. The spectral type we derive for the source is consistent with a $\rm 1.2 \, M_{\odot}$ star inferred from the ALMA observations. Despite its massive circumstellar disk, we find little to no evidence for ongoing accretion on the star. The CO maps reveal a disk that is unusually compact along the vertical direction, consistent with its appearance in scattered light images. The gas disk extends about twice as far away as both the submillimeter continuum and the optical scattered light. CO is detected from two surface layers separated by a midplane region in which CO emission is suppressed, as expected from freeze-out in the cold midplane. We apply a modified version of the Topographically Reconstructed Distribution method presented by \citet{Dutrey2017} to derive the temperature structure of the disk. We find a temperature in the CO-emitting layers and the midplane of $\sim$33 K and $\sim$20 K at $\rm R<200$ au, respectively. Outside of $\rm R>200$ au, the disk's midplane temperature increases to $\sim$30 K, with a nearly vertically isothermal profile. The transition in CO temperature coincides with a dramatic reduction in the sub-micron and sub-millimeter emission from the disk. We interpret this as interstellar UV radiation providing an additional source of heating to the outer part of the disk. 

\end{abstract}

\keywords{circumstellar matter – planets and satellites: formation – protoplanetary disks}


\section{Introduction} \label{sec:intro}

Protoplanetary disks are the birth site of planetary systems, and their structure can provide key insights on the planet formation process itself \citep[][and references therein]{Andrews2020}. The radial structure of protoplanetary disks has been thoroughly studied in both gas and dust components thanks in large part to ALMA \citep[e.g.,][]{Andrews2018}. These observations have revealed the prevalent occurrence of small scale structures such as rings and spiral arms, which have been associated with planets and other physical mechanisms acting in the disks \citep{Dong2016, Christiaens2014, Pinte2019, Zhang2015, Gonzalez2017}. For disks that have been mapped in both continuum and line emission, it has been revealed that the submillimeter-emitting dust component is much less extended radially than the gas component \citep{Mannings1997,Ansdell2018, Boyden2020}. This is generally interpreted as the result of radial drift induced by the drag forces that the gas induces on large dust grains \citep[e.g.,][]{BirnstielAndrews2014, Cleeves2016}, although differences in integrated optical depth could also be at play \citep{Facchini2017}. To fully characterize a protoplanetary disk, it is therefore important to use multiple tracers. In addition to submillimeter continuum and line emission maps, scattered light can be used to trace micron-sized grains, which are expected to be well coupled to the gas \citep[e.g.,][]{Barriere2005} but whose surface brightness is independent of the local temperature.

The vertical structure of protoplanetary disks, on the other hand, has not been studied in as much detail due to a combination of optical depth and projection effects. Disks are expected to be in vertical hydrostatic equilibrium, whereby the vertical component of stellar gravity is balanced by the thermal pressure of the gas. Scattered light images confirm that small dust is lifted up to a flared upper surface, \citep[e.g.,][]{Avenhaus2018} but the height of this $\tau= 1$ surface above the midplane can be any number of scale heights depending on the local surface density. While the mass of the central star, which determines the first force, can be derived kinematically from the gas' orbital motion \citep[e.g.,][]{Huelamo2015, Simon2019}, the pressure gradient is set by the vertical density and temperature profiles in the disk, which is not as readily accessible to observations. 
Interestingly, recent studies have provided model-independent methods to tomographically determine the gas temperature structure of protoplanetary disks \citep{Dutrey2017, Pinte2018, Dullemond2020, Teague2020}.
These methods are applicable to spatially and kinematically resolved disks where the gas component is sufficiently optically thick so that the gas intensity can be transformed into brightness temperature via the Plank equation. These methods open new avenues to address questions related to
vertical snow lines \citep{Pinte2018}, the extent of gas depletion in frozen-out regions \citep{Dullemond2020}, and the role of external sources of irradiation such as the interstellar radiation field \citep{Dutrey2017}.


While these temperature retrieval methods can be used for disks at moderate inclination \citep{Pinte2018, Dullemond2020}, the spatial confusion introduced by projection effects can only be alleviated by kinematics under favorable circumstances. Disks at high inclinations ($i \gtrsim 75$\degr), often dubbed edge-on disks, offer a distinct advantage by preventing the two disk surfaces to be projected on top of one another, making the tomographic temperature reconstruction much easier \citep{Dutrey2017}.
Our team has designed a method to identify new edge-on protoplanetary disks using {\it Spitzer} mapping of star-forming regions and obtaining confirmation images with {\it HST}, identifying a dozen new such systems to complement the existing sample \citep{Stapelfeldt2014}. 

In this study, we focus on SSTC2D\,J163131.2-242627, hereafter Oph163131, a member of the Ophiuchus star-forming region \citep{Evans2009, Hsieh2013} located at a distance of $d =147 \pm 3$~pc \citep{Ortiz-Leon2017}. The central source of Oph163131 does not have a published spectrum, and therefore the stellar parameters of this source were highly uncertain prior to this study. The disk surrounding Oph163131 was unambiguously revealed to be edge-on with a radius of about 190\,au in {\it HST} imaging \citep{Stapelfeldt2014}, as well as in subsequent ALMA continuum survey observations \citep{Cox2017, Cieza2019}. We have obtained dedicated high-resolution ALMA observations, which are presented in \cite{Villenave2020}. In a companion paper (Wolff et al., in prep., hereafter Paper\,I), we simultaneously modeled the {\it HST} scattered light image and the ALMA continuum map. The latter is vertically much thinner, and the radiative transfer model demonstrates that the disk has undergone significant settling.


In this paper, we present and analyze new ALMA CO observations of the Oph13131 disk, as well as the first optical and infrared spectra of this star. In particular, we develop a slightly modified version of the Tomographically Reconstructed Distribution (TRD) method \citep{Dutrey2017} to account for non-perfectly edge-on disk, and we apply it to the disk around Oph163131 in both $^{12}$CO J = 2 -- 1 and $^{13}$CO J = 2 -- 1 isotopologues. Interferometric and spectroscopic observations of Oph163131 are presented in Section \ref{sec:observ}. Observational results such as spectral typing of the central source and dynamical mass estimates are presented in Section \ref{sec:Results}. In Section \ref{sec:model}, we describe the method used to reconstruct the temperature structure of Oph163131 and present its temperature distribution. In Section \ref{sec:discussion}, we put Oph163131 in context by comparing it with other disk sources with independently measured temperature profiles. We contrast the inner disk region's temperature against models that reproduce the ALMA dust continuum and scattered light images. In addition, we present models that include photodissociation, photodesorption, and enhanced UV irradiation as possible mechanisms to explain the observations. In Section \ref{sec:conclusion}, we present our conclusions.

\section{Observations and data reduction} \label{sec:observ}

\subsection{ALMA Continuum and CO Maps}

We observed Oph163131 with ALMA in Band 6 as part of project 2016.1.00771.S (PI: Duch\^ene). The spectral set-up was divided into two 1.9\,GHz continuum spectral windows, of rest frequency 216.5\,GHz and 219.6\,GHz, and three spectral windows that were set-up to include the $^{12}$CO, $^{13}$CO, and C$^{18}$O J = 2 - 1 transitions at 230.538\,GHz, 220.399\,GHz and 219.56\,GHz, respectively. The line spectral windows have native velocity resolutions of 0.16\,km\,s$^{-1}$ for $^{13}$CO, and C$^{18}$O and 0.085\,km\,s$^{-1}$ for $^{12}$CO. The source was observed using two different array configurations, specifically a compact and an extended configuration on April 25, 2017 and on July 7, 2017, respectively. The baselines ranged from 15\,m to 2650\,m between the two configurations, and the total observing time on source was 40 minutes. The raw data were calibrated using the CASA pipeline version 4.7.2.

To maximize the dynamical range, we performed phase self-calibration from the continuum data from both configurations. We extracted the continuum images using the CASA \texttt{tclean} task on the combined data, with a Briggs parameter of 0.5. The resulting beam size is 0\farcs23$\times$0\farcs13. The continuum image is further analyzed in Paper\,I. 
After applying the continuum self-calibration solutions to all spectral windows, we extracted the emission lines from the calibrated visibilities by subtracting continuum emission using the \texttt{uvcontsub} function in CASA. In addition to the continuum subtracted data, we produced a non-continuum subtracted dataset to test whether our modeling method is affected by possible over-subtraction of the continuum in optically thick regions of the disk. We binned the native spectral resolution to 0.25\,km\,s$^{-1}$ and used Briggs parameter of 0.5 to create higher signal-to-noise (S/N) line images. The resulting beam were 0\farcs25$\times$0\farcs20 for $^{12}$CO and 0\farcs26$\times$0\farcs20 for $^{13}$CO, and C$^{18}$O. The S/N of the C$^{18}$O isotopologue was not high enough to produce a dataset we can analyze.

\subsection{JCMT CO map} \label{sec:JCMT_obs}

On April 16, 2011, we obtained $^{12}$CO J = 3 - 2 data from the Heterodyne Array Receiver Program (HARP) at the 15-m James Clerk Maxwell Telescope (JCMT). An  8\arcmin$\times$8\arcmin \, map around Oph\,163131 was acquired at a rest frequency of 345 GHz using 8192 ACSIS channels with a bandwidth coverage of 250 MHz at a spectral resolution of 30 kHz (0.025 $\rm km \, s^{-1}$). The observations were performed in good weather conditions with measured zenith opacities at 225 GHz ($\uptau_{\rm 225GHz}$) varying from 0.09 to 0.1. 
 
The HARP data was reduced using the \texttt{SMURF} code and a data cube was created with the \texttt{MAKECUBE} task. The pixel size of the data cube was set to 6\arcsec \, which oversamples the 14\arcsec \, JCMT main beam size (at 345 GHz).
In order to increase the S/N of the data, the native velocity resolution was binned to 0.25 $\rm km \, s^{-1}$. At this velocity resolution the measured antenna temperature noise is $\rm \sigma_{TA*} = 0.3 \, K$ per 0.25 $\rm km \, s^{-1}$, or equivalently $\rm \sigma_{Jy} = 9 \, Jy$ per 0.25 $\rm km \, s^{-1}$. This conversion was performed using $\rm S(Jy)=15.6 \, T_{A}(K)^*/\eta_ a$, where $\eta_a=0.52$ is the aperture efficiency of HARP\footnote{\url{http://www.eaobservatory.org/jcmt/instrumentation/heterodyne/calibration/}}. 

\subsection{SOAR Optical Spectroscopy}

We obtained an optical spectrum of Oph163131 on May 19, 2013, with the Goodman spectrograph installed on the SOAR 4.1m telescope at Cerro Tololo. Two 900 second spectra were acquired while dithering along the 0\farcs46 slit. The object was observed at an airmass of 1.01. We used the 400 lines/mm grating and the red camera, which yields a spectral resolution of $\sim$1850 over the full 3400--7500~\AA \, wavelength range. Spectral images were reduced using the IRAF \texttt{ccdproc} package. A narrow feature in the quartz flat field lamp was removed using procedures within the IRAF \texttt{twodspec} package, as customized by SOAR observatory staff (``Removing arc from GOODMAN 600b flatfield", T. Ribeiro, private communication).
Spectral extraction and wavelength calibration referencing SOAR mercury/argon arcs were performed using IRAF \texttt{doslit} procedures.

\subsection{IRTF Near-infrared Spectroscopy}

Our near-infrared (NIR) observations of Oph163131 were carried out with the 3.0 m NASA Infrared Telescope Facility (IRTF) on Maunakea, Hawaii on 2019 May 16. We used the upgraded SpeX instrument in the short crossdispersed (SXD) mode, which covers the 0.7 $\micron$ to 2.5 $\micron$ region in each exposure \citep{Rayner2003}, and the 0\farcs5 wide slit, which gives a resolving power of $\rm R = 1\,200$. Our target was nodded along the slit, with two exposures taken at each nod position. The total exposure time was driven by our goal to get a spectrum with signal-to-noise ratio S/N $>30$ in the \textit{H}-band. The individual exposure times were limited to three minutes to ensure that the telluric emission lines would cancel when consecutive images taken at alternate nod positions were subtracted. After observing Oph163131, we acquired an A0 telluric standard star within 0.1 air masses for telluric correction purposes.

A thorium-argon lamp was observed for wavelength calibration, and a quartz lamp spectrum was acquired for flat fielding. 
The SpeX data were flat-fielded, extracted, and wavelength-calibrated using Spextool \citep{Cushing2004}. After extraction and wavelength calibration, the individual extracted spectra were co-added using \texttt{xcombspec}. \texttt{Xtellcor} was then used to construct a telluric correction model using the observed A0 standard, after which the observed spectrum of the target was divided by the telluric model. Finally, \texttt{xmergeorders} was used to combine the spectra in the separate orders into one continuous spectrum.

Although the SXD mode of SpeX provides a wavelength coverage down to 0.7~$\micron$, the low S/N at wavelengths shortward of about 1~$\micron$ precludes connecting the NIR spectrum of Oph163131 with its optical counterpart.

\section{Observational Results}\label{sec:Results}

\subsection{Spectral Characterization}
\label{subsec:sepctral_characterization}
To the best of our knowledge, the optical and near-infrared spectra of Oph\,163131 presented here are the first of this source. In edge-on disks, the smooth scattering properties of astronomical dust preserve the spectral features associated with the central source \citep{Appenzeller2005}. Therefore, we use our SOAR and SpeX spectra to 1) estimate the spectral type, hence the physical properties of the central star, and 2) evaluate the accretion status of the system based on the strength of its emission lines.

\begin{figure*}
\epsscale{1.1}
\plottwo{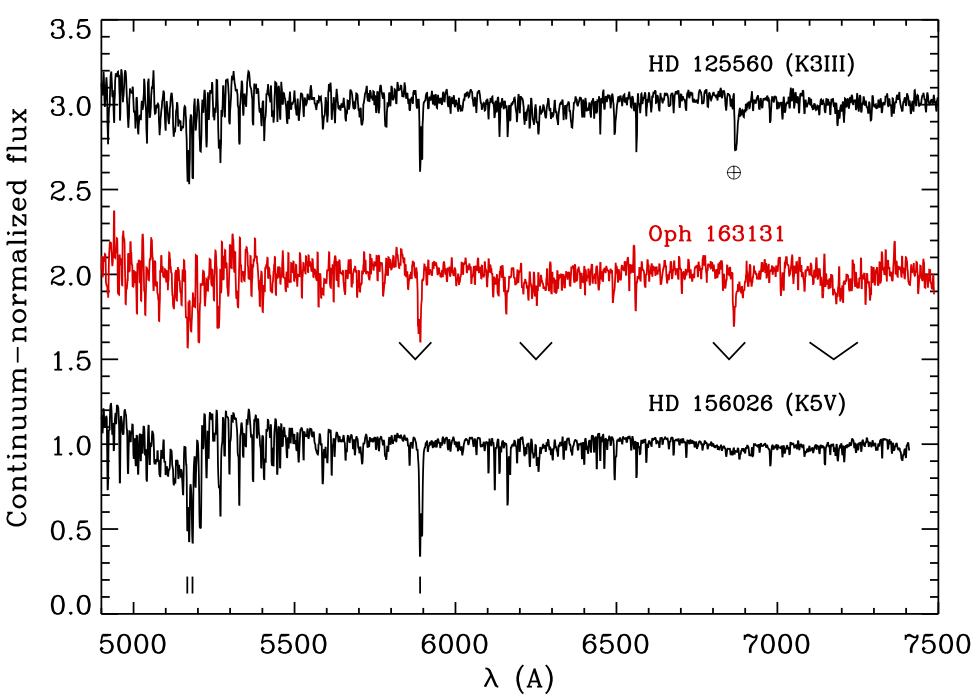}{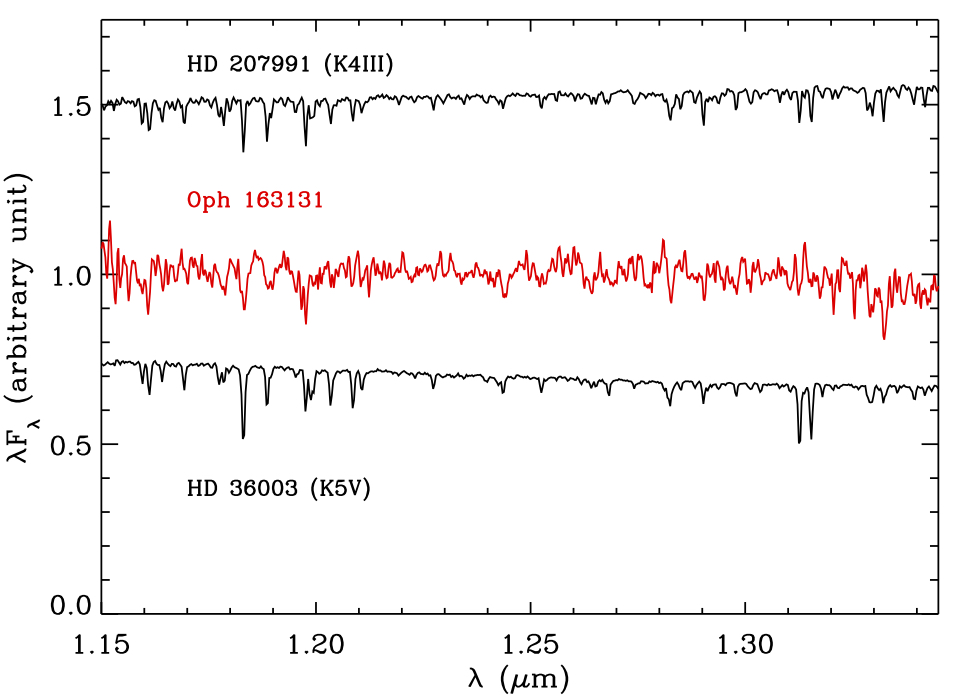}
\plottwo{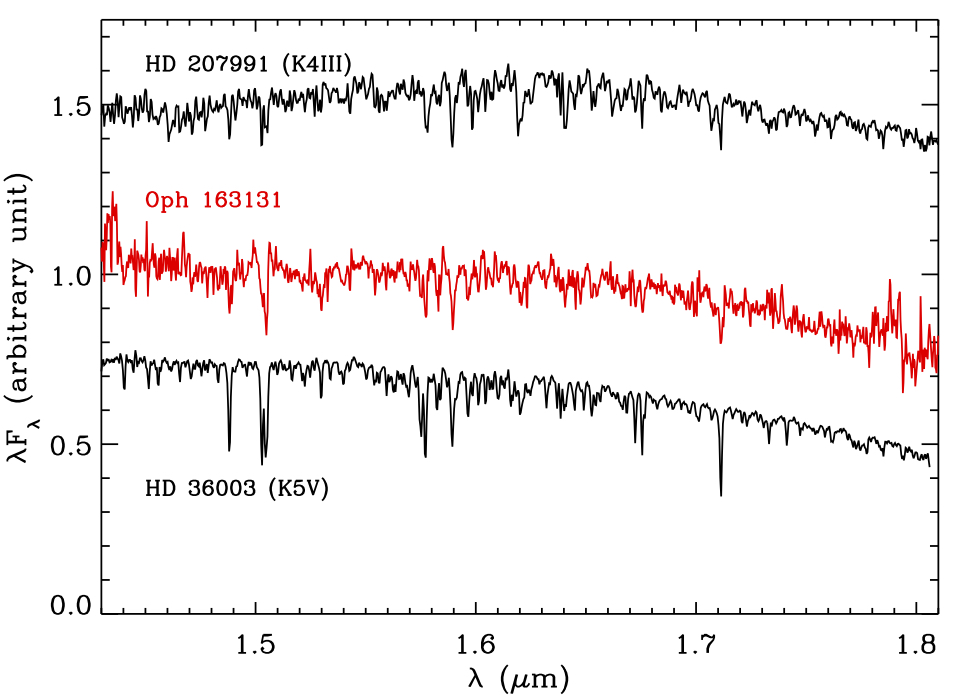}{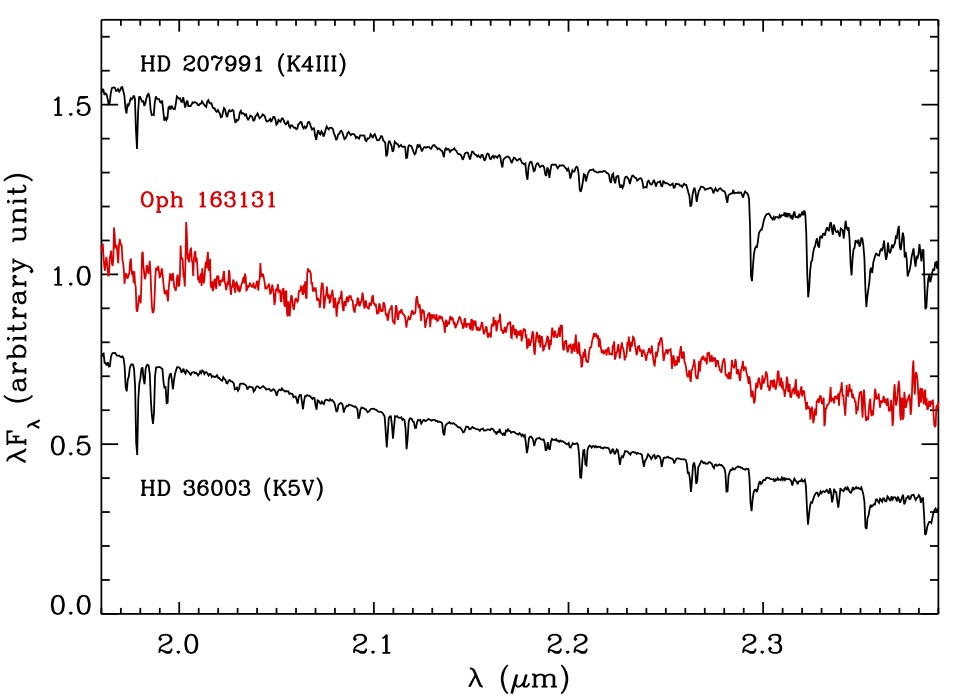}
\caption{Spectrum of Oph\,163131 in the optical (top left panel) and near-infrared (other panels) compared to dwarfs and giants from the MILES, STELIB and SpeX libraries. In the top left panel, a known telluric feature, weak TiO bandheads, and Mg and Na features are labeled under the top, middle, and bottom spectra, respectively. The spectra are offset vertically for clarity. The optical spectrum is normalized to a low-order polynomial fit to the continuum whereas the near-infrared spectra are spectrophotometrically calibrated.}
\label{fig:opt_spectrum}
\end{figure*}

In the absence of spectrophotometric calibration and since scattering off the disk can alter the apparent colors of the central source for our optical spectrum, we focus our spectral analysis on narrow spectral features rather than on the global shape of the continuum. After extracting the 4700--7500\,\AA\ spectral range, where the signal-to-noise is sufficient, we fit the spectrum with a polynomial of order 6 and divide it by spectrum. Polynomial orders lower than 4 or higher than 7 leave small unwanted residuals, which do not strongly impact the stellar classification but look artificial in the stellar spectrum. The resulting continuum-normalized spectrum is shown in the top-left panel of Figure\,\ref{fig:opt_spectrum}. The most prominent feature in the spectrum is the sharp Na\,I D absorption line at 5890\,\AA\ and the complex of lines surrounding the 5150\,\AA\ Mg line. The main TiO bands (at around 5900, 6200, 6800, and 7100 \AA), which are typically seen in K5 and later type stars, are weak in the spectrum of Oph 163131. The combination of these observations suggests that, in the optical, Oph163131 is an early- to mid-K spectral type source \citep{Jacoby1984,Allen1995}. In the near-infrared, the spectrum of Oph\,163131 (also shown in Figure\,\ref{fig:opt_spectrum}) shows a number of atomic features across the $J$, $H$ and $K$ band, as well as modest CO bandheads beyond 2.29\,$\mu$m. The relative strengths of these features, when compared to the spectral library of \cite{Rayner2009}, as well as the overall shape of the continuum, are consistent with a mid to late-K type star.

We further this analysis by comparing the spectrum of Oph\,163131 to that of templates from the MILES and STELIB \citep[in the optical,][]{Sanchez2006,Falcon2011,LeBorgne2003} and IRTF/SpeX \citep[in the near-infrared][]{Rayner2009} libraries. We perform the comparison with both dwarfs and giants\footnote{Unfortunately, neither library contains spectra of subgiants}. To circumvent issues such as stellar rotation and line veiling induced by accretion, we limit our analysis to a visual inspection that focuses equally on the strength of individual absorption features and on small-scale continuum structure related to molecular bands. This is sufficient for the level of precision needed for our analysis. The closest match for the optical spectrum of Oph\,163131 is a K3--K4 giant, while in the infrared the spectrum resembles a K4--K5 dwarf. Such discrepant spectral type measurement at different wavelengths could be simply due to the visual characterization of the stellar spectra. However, it might also be a real phenomenon, as spectroscopic studies at high-resolution have shown that starspots on the surfaces of young stars produce an observed temperature dependency with wavelength \citep[see e.g.,][]{Flores2020, Sokal2018}. The stellar luminosity (and hence the luminosity classification) of Oph163131 is uncertain because of dilution by an unknown amount through scattering off the disk. Therefore, it cannot be calculated using the distance to the star and photometry. In any case, since our best match luminosity type lay between dwarfs (V) and giants (III),  we adopted a sub-giant (IV) luminosity classification for Oph 163131. This is consistent with previous work where sub-giant surface gravities have been measured for young stars \citep{Doppmann2005, Flores2019}. In summary, and considering the optical and IR spectra of the star, we classify Oph163131 as a K4\,IV star, with a spectral type uncertainty of about one subtype. This implies a stellar mass of $\sim1\, \rm M_\odot$ based on recent dynamical mass measurements for stars in Taurus and Ophiuchus \citep{Simon2019} and also from stellar evolutionary models \cite[e.g.,][]{Feiden2016,Somers2015}.

\begin{figure}
\epsscale{1.1}
\plotone{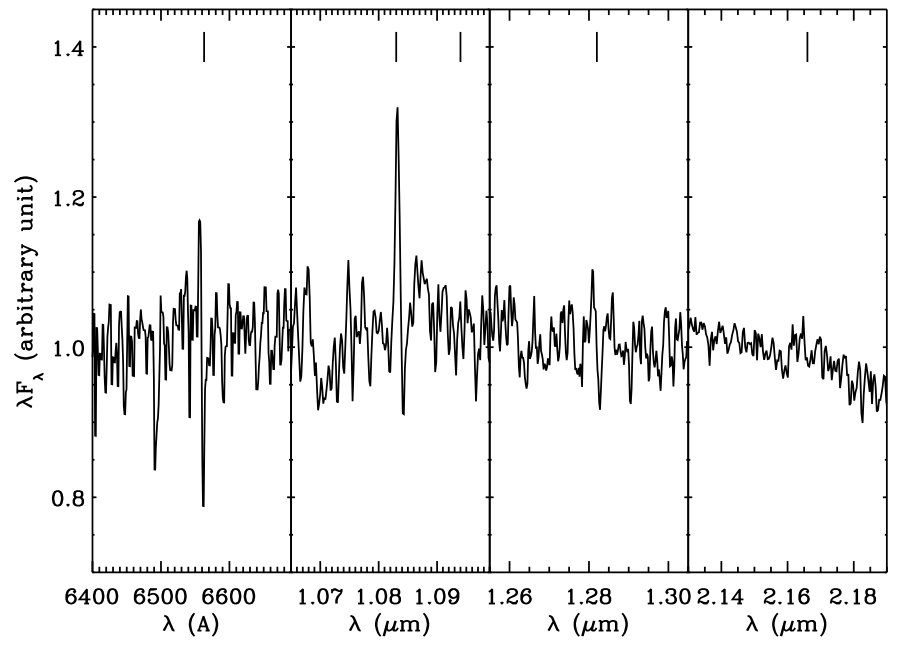}
\caption{Main emission lines in the spectrum of Oph\,163131. From left to right, the lines are H$\alpha$, He\,I 1.08\,$\mu$m and H Pa\,$\gamma$, H Pa\,$\beta$ and H Br\,$\gamma$. The wavelength of each feature is indicated by a vertical bar above the spectrum. \label{fig:emission_lines}}
\end{figure}

The optical spectrum of Oph\,163131 stands out from that of other edge-on disks systems for the absence of strong emission lines \citep[e.g.,][]{Appenzeller2005}. Indeed, we only detect a very weak H$\alpha$ emission component, with an equivalent width of 0.6$\pm$0.1\,\AA, superimposed on a significant photospheric absorption feature. Such a weak H$\alpha$ line, especially in the absence of other accretion- or outflow-related emission lines, generally indicates that the central source is currently not accreting at an appreciable level \citep[e.g.,][]{Barrado2003, Duchene2017}, i.e., Oph163131 can be classified as a Weak-line T Tauri star. This assessment is confirmed at first glance by our near-infrared spectrum, in which no significant Pa\,$\gamma$, Pa\,$\beta$ or Br\,$\gamma$ emission is detected (see Fig. \ref{fig:emission_lines}), although it is worth pointing out that these lines are less sensitive to low levels of accretion onto T\,Tauri stars \citep[e.g.][]{Folha2001,Alcala2014}. 

\begin{figure}
\epsscale{1.1}
\plotone{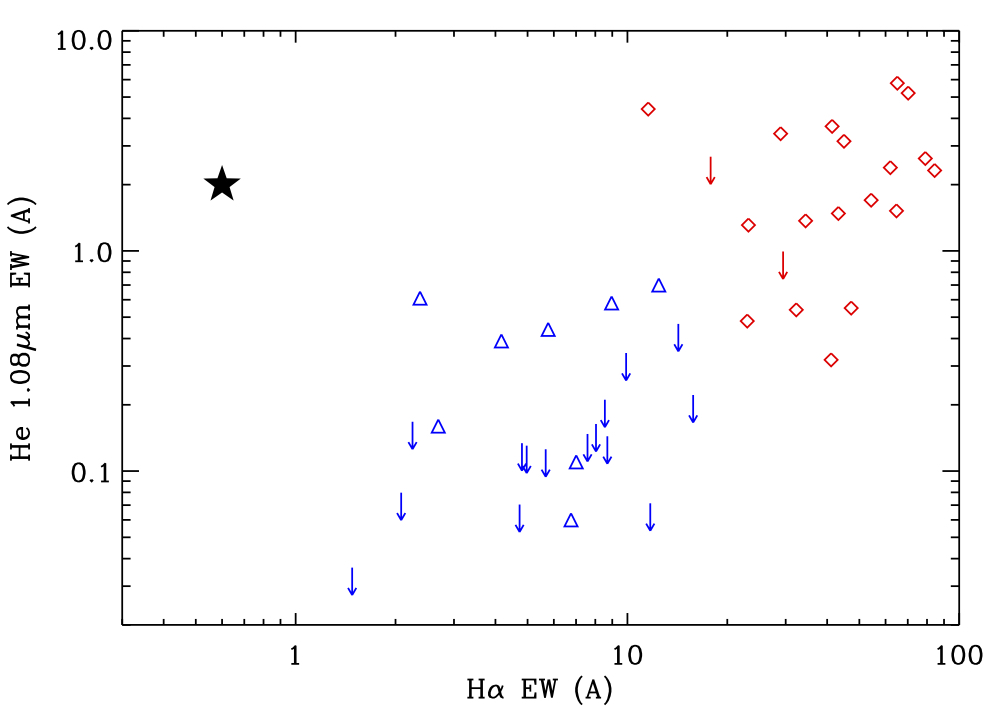}
\caption{Comparison between the equivalent width of the H$\alpha$ and He\,I\,1.08\,$\mu$m lines among accreting (red diamonds) and non-accreting (blue triangles) T\,Tauri stars. Data are from \cite{Edwards2006}, \cite{Manara2013} and \cite{Alcala2014}. Arrows mark upper limits on the He\,I line. The black star indicates results for Oph163131.}
\label{fig:ha_he}
\end{figure}

The only other significant emission line detected in our spectra is the 1.08\,$\mu$m He\,I line, with an equivalent width of 2.0$\pm$0.2\,\AA. This line has a complex origin, with possible contributions from chromospheric activity, disk and/or stellar wind, and accretion onto the central star \citep{Edwards2006}. Surprisingly, all T\,Tauri stars with an He\,I emission line as strong as Oph163131 are accretors (see Fig.\,\ref{fig:ha_he}), leading to an apparent paradox. The ultimate diagnostic of the origin of the He\,I line lies in the line profile, \citep{Fischer2008,Thanathibodee2018}, which our low-resolution data do not allow us to analyze. We are therefore left with an ambiguous classification for Oph163131. We note that the edge-on nature of the object could lead to biased estimates of line equivalent width if the emission is spatially distinct from the star itself. However, if the emission is chromospheric in nature, its origin is at the stellar surface, and thus our equivalent width estimates would be unbiased. 

In summary, while Oph163131 is unlikely to be accreting at a high rate, existing data are consistent with both 1) an object with a modest amount of accretion (and a remarkably weak H$\alpha$ line), and 2) a non-accreting, chromosphere-dominated emission line spectrum (albeit with an unusually strong He\,I\,1.08\,$\mu$m line). Only higher spectral resolution observations can solve this ambiguity.


\subsection{Dynamical Mass Measurement}
\label{sec:dynamical_mass}

\begin{figure*}[ht!]
\centering
\plotone{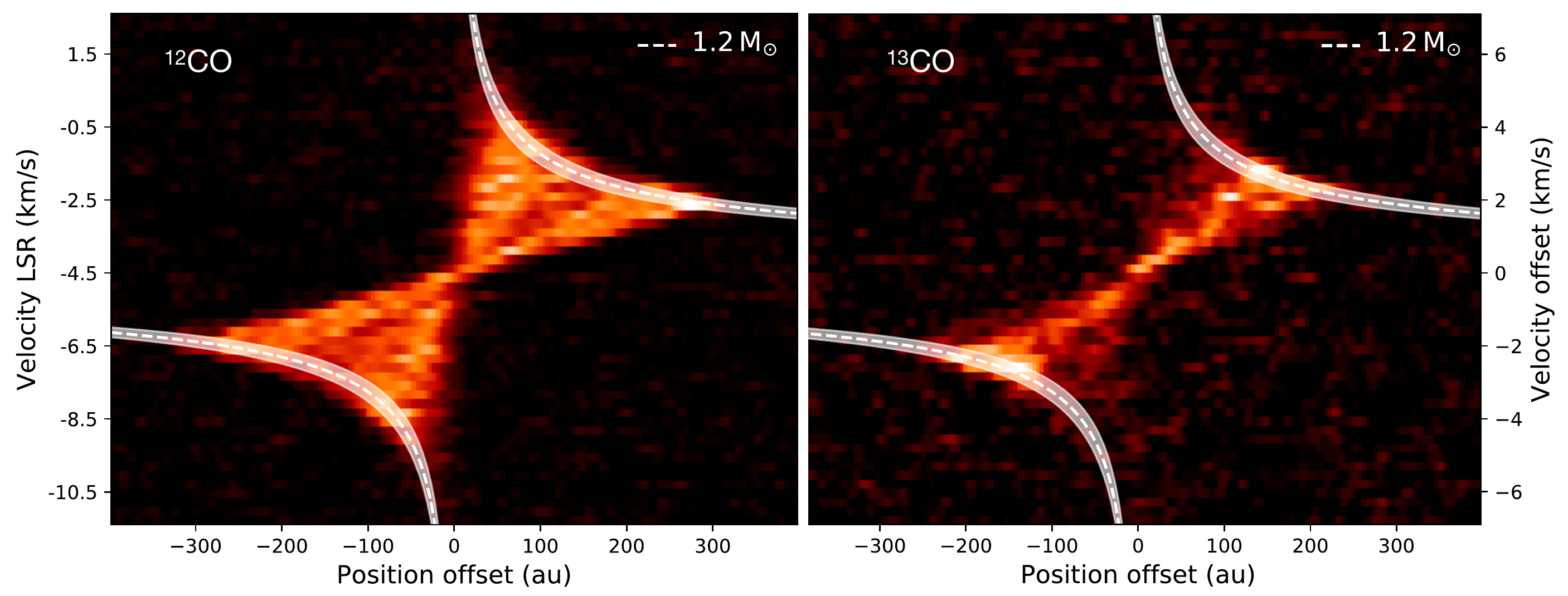}
\caption{Position-velocity diagrams along the major axis of Oph163131. The $^{12}$CO(2--1) isotopologue is shown on the left panel with the LSR velocity of the system in left y-axis. The $^{13}$CO(2--1) is shown on the right panel using the velocity offset coordinates and assuming a systemic velocity of $v_{sys}=-4.4 \rm \, km \, s^{-1}$. The white dashed lines correspond to a Keplerian velocity profile for a star of 1.2 $\rm M_{\odot}$. The white filled region is bracketed by upper and lower limits of 1.4 $\rm M_{\odot}$ and 1.0 $\rm M_{\odot}$, respectively.}
\label{fig:dynamical_masses}
\end{figure*}

In the previous section, we used the spectral type derived for Oph163131 and results from stellar evolutionary models to infer the mass of the star. However, it is possible to directly measure the stellar mass using the disk's CO data and the `position-velocity (PV) diagram' technique \citep[see for example,][]{Huelamo2015}. 

In general, dynamical mass measurements using this technique work better when the emission from a disk is optically thin (e.g., $^{13}$CO or C$^{18}$O). This is because optically thin molecules trace the emission closer to the disk's midplane, or equivalently because there is less confusion with emission arising high up in the disk's surface. In the case of highly inclined disks $i \gtrapprox 85$\degr, however, the position of the disk's major axis  approximately coincides with the disk midplane. Therefore low-elevation emission can be observed even for optically thick tracers such as $^{12}$CO. Since Oph163131 is known to host an almost perfectly edge-on disk, we can confidently calculate the dynamical mass of the star from either of the CO tracers. 

The computed PV diagrams in $^{12}$CO and $^{13}$CO along the major axis of the disk are shown in Figure \ref{fig:dynamical_masses}. In each case, Keplerian velocity profiles of an inclined disk with $i=87\degr$ (the best-fit inclination to the ALMA continuum map from Paper\,I) with a stellar mass of  $\rm 1.2 \, M_{\odot}$ were overlaid. The systemic velocity of the disk was estimated to be $v_{sys}=-4.4$~$\rm km \, s^{-1}$, which is further confirmed in Section \ref{sec:disk_morphology} using $\rm ^{12}CO$ integrated spectrum and channel maps. Although the Keplerian profiles were matched to the data by eye, we are confident in this technique as we have tested it with synthetic disk models computed with MCFOST \citep{Pinte2006,Pinte2009}. We assign conservative upper and lower mass limits to Oph163131 of 1.4 $\rm M_{\odot}$ and 1.0 $\rm M_{\odot}$, respectively. We conclude that the dynamical mass measured for Oph163131 is in agreement with the spectral type derived in Section \ref{subsec:sepctral_characterization}. 


\subsection{Cloud contamination}
A crucial part of the temperature reconstruction method explained in Section \ref{sec:model} requires us to determine whether there is emission on a scale much larger than the disk, for example, caused by a molecular cloud at the position of Oph163131. To investigate whether this large scale emission is present or not, we use $^{12}$CO~J~=~3--2 single-dish observations acquired with HARP/JCMT (see Section \ref{sec:JCMT_obs}). 

\begin{figure*}[ht!]
\centering
\epsscale{1.1}
\plotone{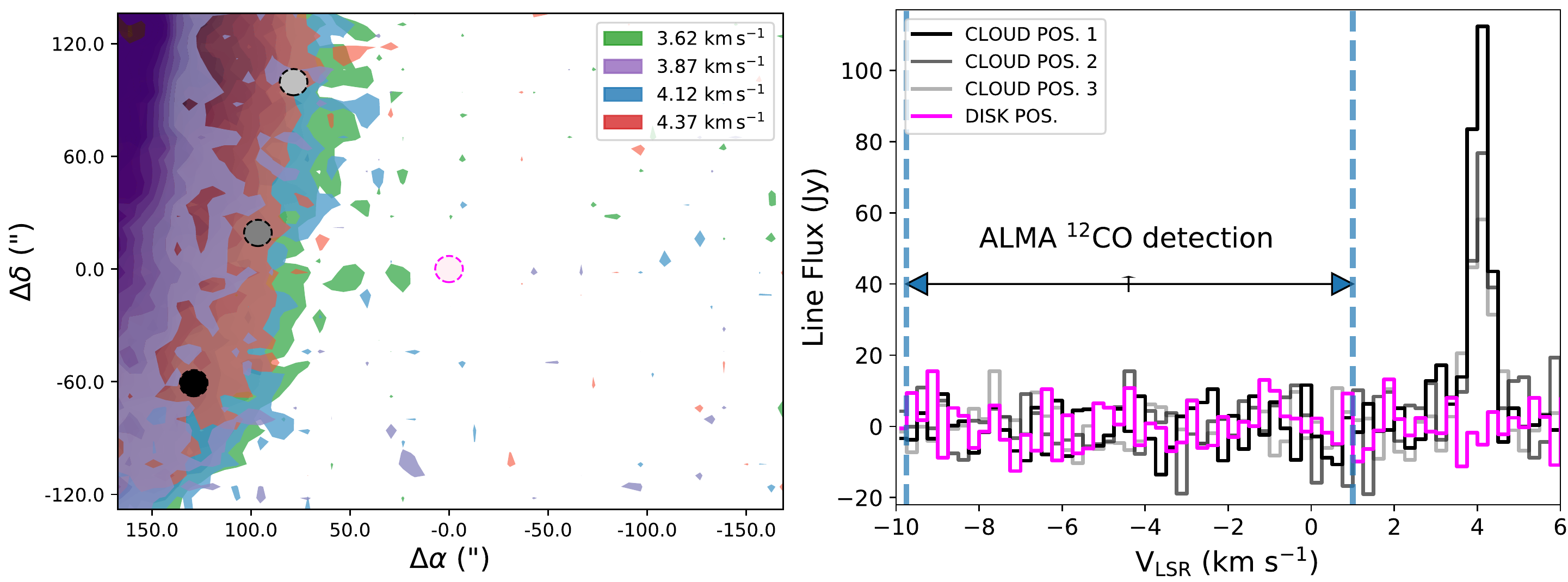}
\caption{The left panel shows the JCMT $^{12}$CO J = 3 -- 2 intensity map of the large scale structure around Oph163131. In colors we displayed four different velocities channels where the emission of the cloud is the strongest (with $v_{LSR}$ ranging from 3.6 to 4.4 $\rm km s^{-1}$). Four $14\arcsec$-diameter (i.e., beam-sized) circular apertures are overlaid at locations within the cloud (light gray, gray, and black) and at the location of Oph163131 (pink). The right panel shows the JCMT spectra obtained at the locations mentioned above, and using the same colors as the apertures in the left panel. The dashed blue vertical lines and the horizontal arrow depicts the velocity range over which the Oph163131 disk is detected above 3$\sigma$ in the $^{12}$CO J = 2 -- 1 ALMA data. There is no spatial or spectral overlap between the Oph163131 disk and the cloud.}
\label{fig:JCMT}
\end{figure*}

\begin{figure}[ht!]
\centering
\epsscale{1.1}
\plotone{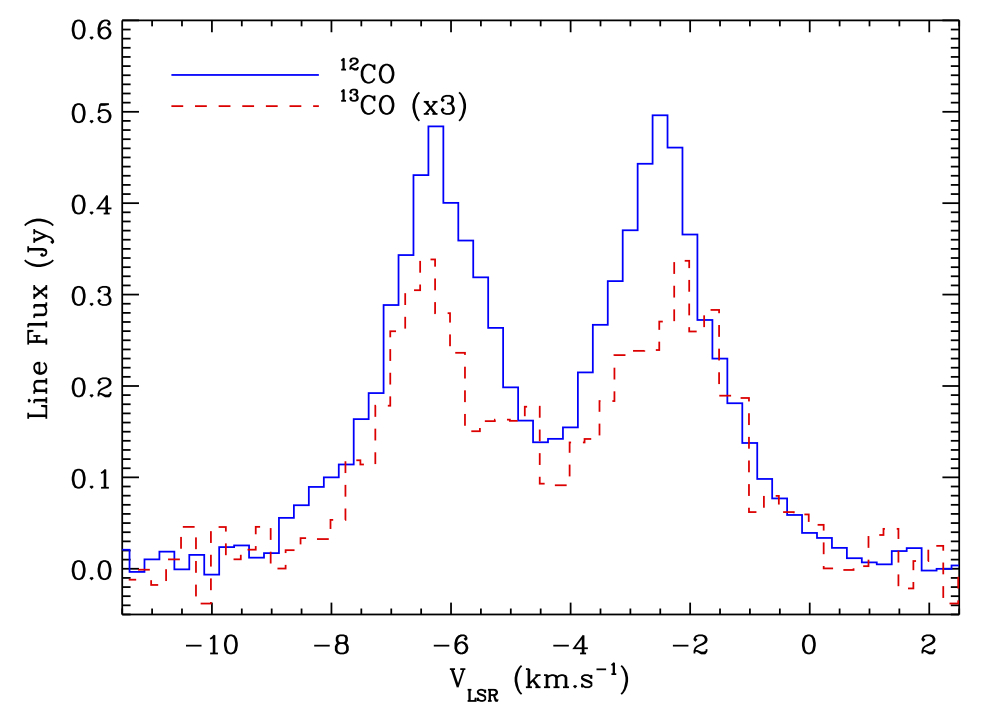}
\caption{ALMA $^{12}$CO (in solid blue) and $^{13}$CO (in dashed red) spectra of the disk around Oph163131. The $^{13}$CO emission was scaled up by a factor of three for illustrative purposes.}
\label{fig:ALMA_spectra}
\end{figure}

A $\sim$6\arcmin$\times$6\arcmin \, JCMT CO map centered at the position of Oph163131 is shown in Figure \ref{fig:JCMT}. Part of the large L1689 molecular cloud is detected East of Oph163131 at velocities ranging from  3.3 to 4.6 $\rm  km \, s ^{-1}$. The spectrum measured at the location of Oph163131 shows no significant emission above 2$\sigma$ in any of the velocity channels. Conversely, the three spectra measured in different locations of the molecular cloud show prominent emission at $\sim4 \,\rm  km \, s ^{-1}$. This velocity range does not overlap with the CO emission from the Oph163131 disk in our ALMA data (see Figure \ref{fig:ALMA_spectra}).
Therefore, we conclude from the JCMT data that the cloud emission does not coincide in the spatial or velocity dimensions with that of the disk.

To further investigate whether low levels of cloud emission could affect the disk emission while being undetected in our JCMT data, we consider the $^{12}$CO and $^{13}$CO ALMA spectra of Oph163131 (Figure \ref{fig:ALMA_spectra}).
The spectra were computed by summing the flux in a rectangular region of $5.4\arcsec$ along the major axis and $1.5\arcsec$ along the minor axis of the disk. The $^{12}$CO emission is detected above a $3\sigma$ level between -9.75 and +1.0 $\rm  km \, s ^{-1}$, while the $^{13}$CO emission is detected above a $3\sigma$ level in the slightly narrower range from -8.0 to -0.75 $\rm  km \, s ^{-1}$. This difference is likely due to the lower S/N detection of the latter isotopologue. 

Both spectra show the characteristic double peak profile seen in other disks \citep[e.g.,][]{Beckwith1993, Mannings1997}. The nearly perfectly symmetric profile displayed in $^{12}$CO is a strong indicator that no emission is added or subtracted to the disk. In cases where the line emission from a disk is contaminated by either a background or foreground cloud 
the observed spectra always display a strongly asymmetric double profile in optically thick tracers such as $^{12}$CO \citep[see for example,][]{Guilloteau2016,Perez2015}.

\subsection{Disk morphology}
\label{sec:disk_morphology}

\begin{figure*}[ht!]
\plotone{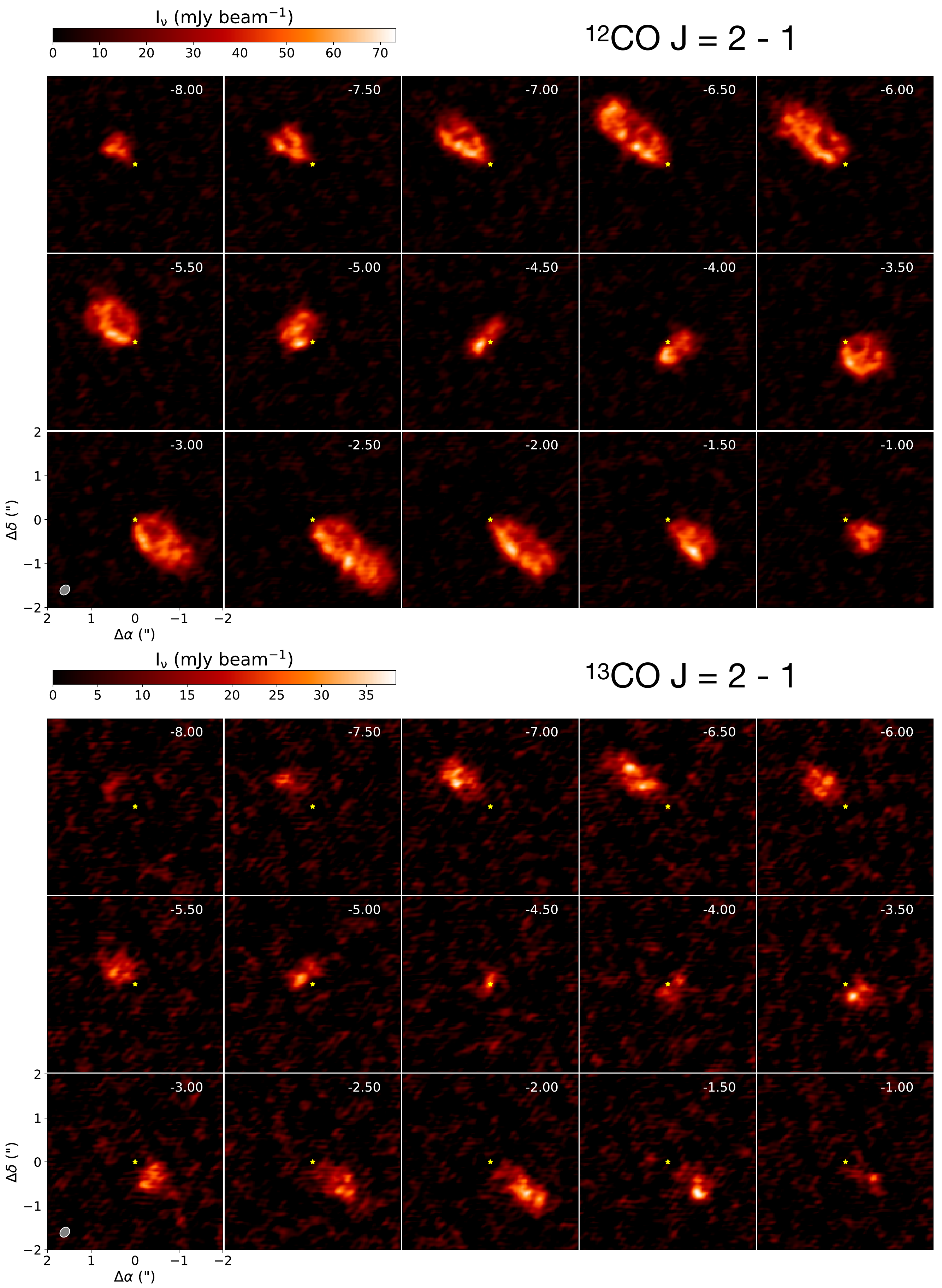}
\caption{$^{12}$CO (top) and $^{13}$CO (bottom) ALMA channel maps of the Oph163131 disk. Each channel shows a square region of $4\arcsec$ in size centered on the predicted stellar position marked with a yellow star. The LSR velocity (in $\rm  km \, s ^{-1}$) is indicated in the top-right of each channel. The synthesized beam 
is shown as a grey ellipse in the lower-leftmost channel.}
\label{fig:ALMA_12CO}
\end{figure*}

We now describe the morphology of the Oph163131 disk as seen by ALMA.  Figure \ref{fig:ALMA_12CO} shows the $^{12}$CO J~=~2~-~1  and $^{13}$CO~J~=~2~-~1 channel maps of the disk. In both isotopologues, Oph163131 exhibits the typical Keplerian velocity pattern of a highly inclined disk \citep[see for example][]{Dutrey2017}, with $^{12}$CO having a stronger detection due to the higher peak signal-to-noise ratio of S/N~$\sim 24$ per 0.25 $\rm  km \, s ^{-1}$ compared to the peak signal-to-noise ratio for $^{13}$CO of only S/N~$ \sim 12$ per 0.25 $\rm  km \, s ^{-1}$. Oph163131 is well spatially resolved in both isotopologues with the $^{12}$CO extending $4.6\arcsec$ along the major axis and $1.2\arcsec$ in the minor axis direction, and $^{13}$CO extending  $3.6\arcsec$ and $0.7\arcsec$ along the major and minor axes, respectively.

\begin{figure*}[ht!]
\epsscale{1.1}
\plottwo{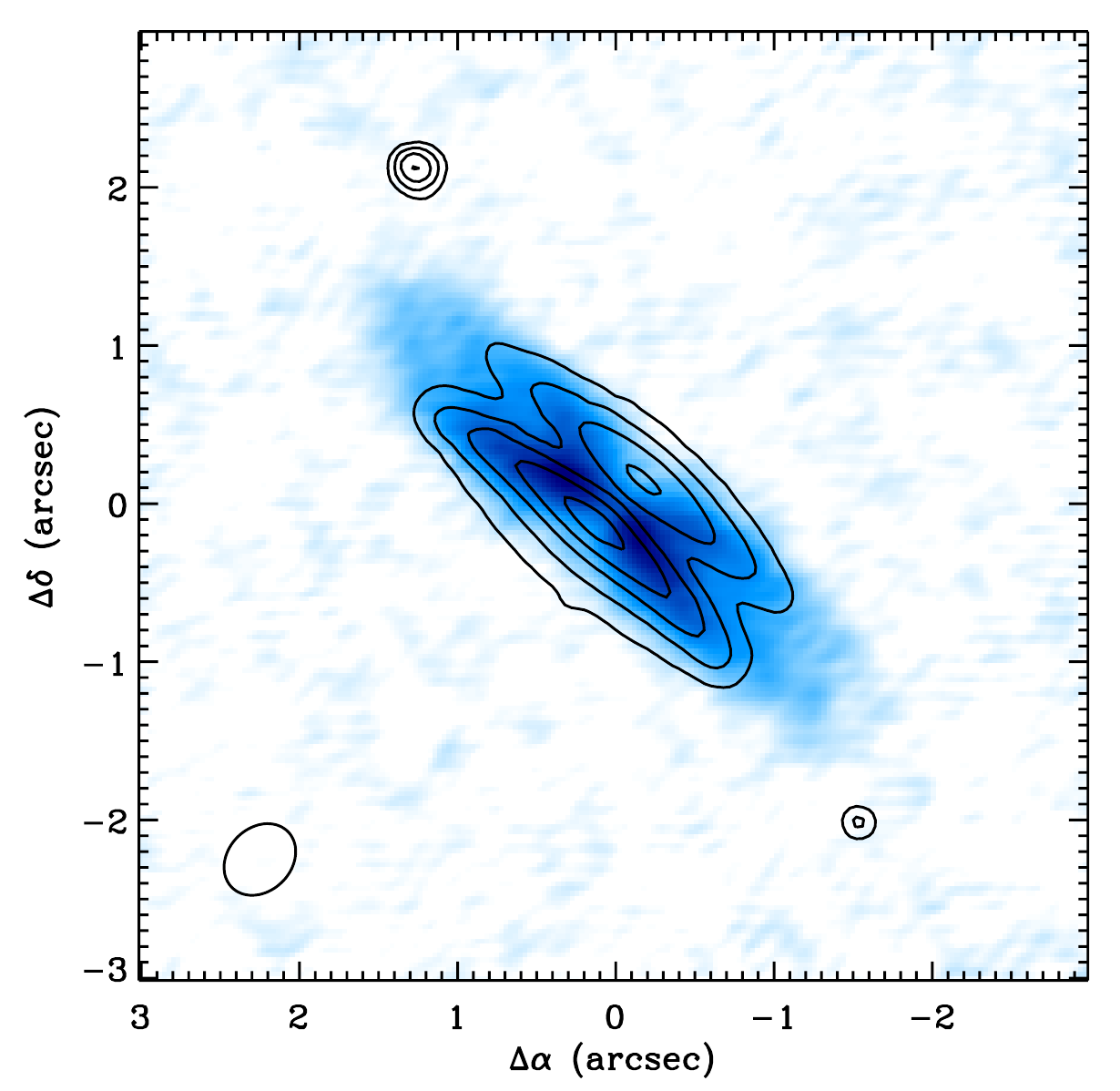}{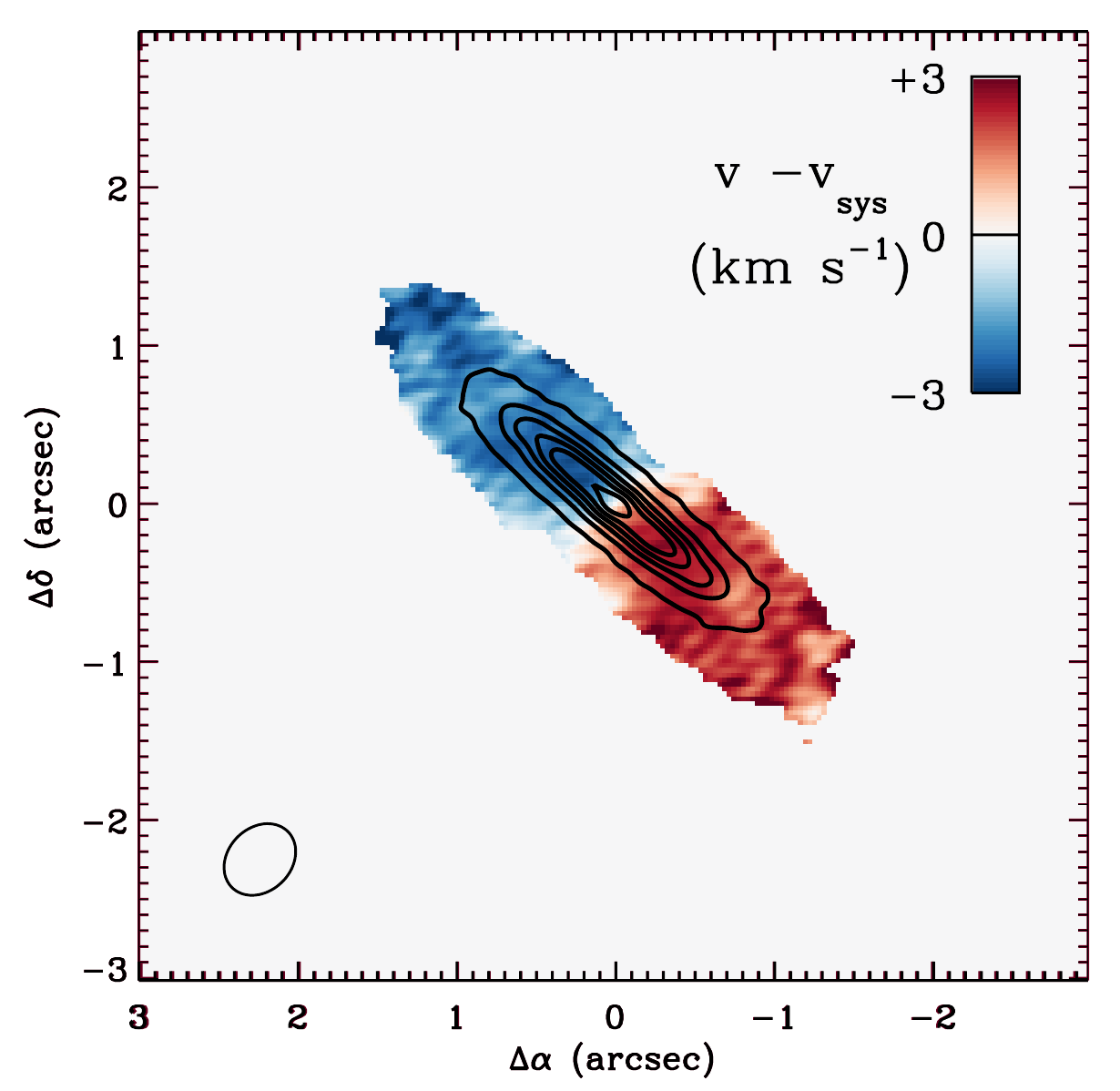}
\caption{Left panel: Overlay of the $^{12}$CO zero moment map shown in colors and the HST F606W image described in Paper\,I in contours of 3, 10, 20, 40 and 60\,$\sigma$. The effective ALMA beam is indicated in the bottom left corner whereas two unrelated stars in the HST image are representative of the point spread function in the optical. Right panel: Overlay of the $^{12}$CO first moment map and the 1.3\,mm continuum map described in Paper\,I. Contours for the continuum map are shown at the 5, 50, 100, 150, 200, 250 and 300$\sigma$ level; the first moment map is only shown where the moment zero exceeds the 5$\sigma$ level.}
\label{fig:HST_ALMA_cont}
\end{figure*}

In $^{12}$CO, the disk displays two clear bright layers along the minor axis (in the Southeast and Northwest directions) separated by a darker middle lane. The projected distance between the two bright layers is $\sim0.45\arcsec$ and the SE layer is $10-20$~mJy brighter than the NW layer. Other ALMA results where CO maps show this type of brightness asymmetry are the Flying Saucer (2MASS J16281370-2431391) \citep{Dutrey2017} and IM Lup \citep{Pinte2018}. The physical explanation for this phenomenon is the viewing angle of the observer with respect to the disk. Oph163131 is tilted slightly away from a perfectly edge-on configuration, with the SE side being the `top' side as seen from the observer. Geometrical and optical depth effects, in particular the thickness of the CO emitting layer and absorption by the cooler midplane layer, result in the `bottom' layer being partially hidden whereas we can directly see a much larger surface area of the `top' layer.  Interestingly, scattered light images of Oph163131 also show the SE layer being brighter than the NW layer, as can be see from Figure \ref{fig:HST_ALMA_cont}, where the $\sim 0.6 \, \mu \rm m$ HST F606W scattered light image of Oph163131 is overlaid on the $^{12}$CO integrated intensity (moment zero) map (see also Paper\,I).

Although the $^{13}$CO emission is resolved along the minor axis of the disk (over-sampled 2.8 times), it is not possible to distinguish the two emitting layers seen in$^{12}$CO. The combination of a lower S/N and smaller vertical size of the disk in $^{13}$CO, prevent us from distinguishing any additional structure in this isotopologue.

\begin{figure*}[ht!]
\plottwo{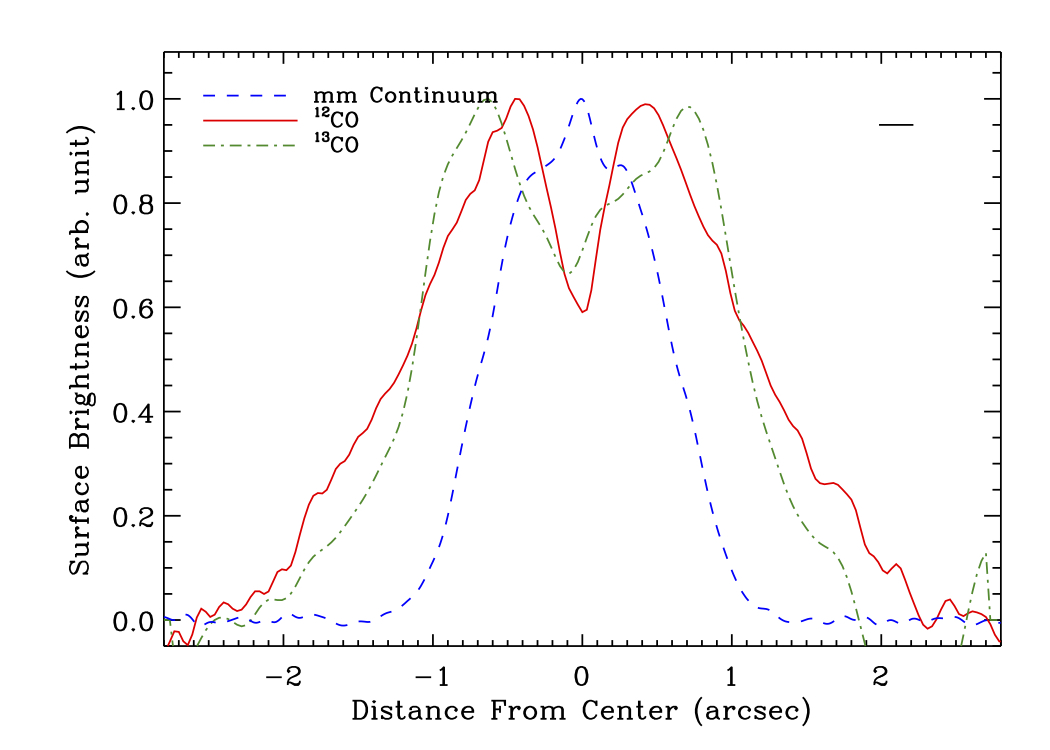}{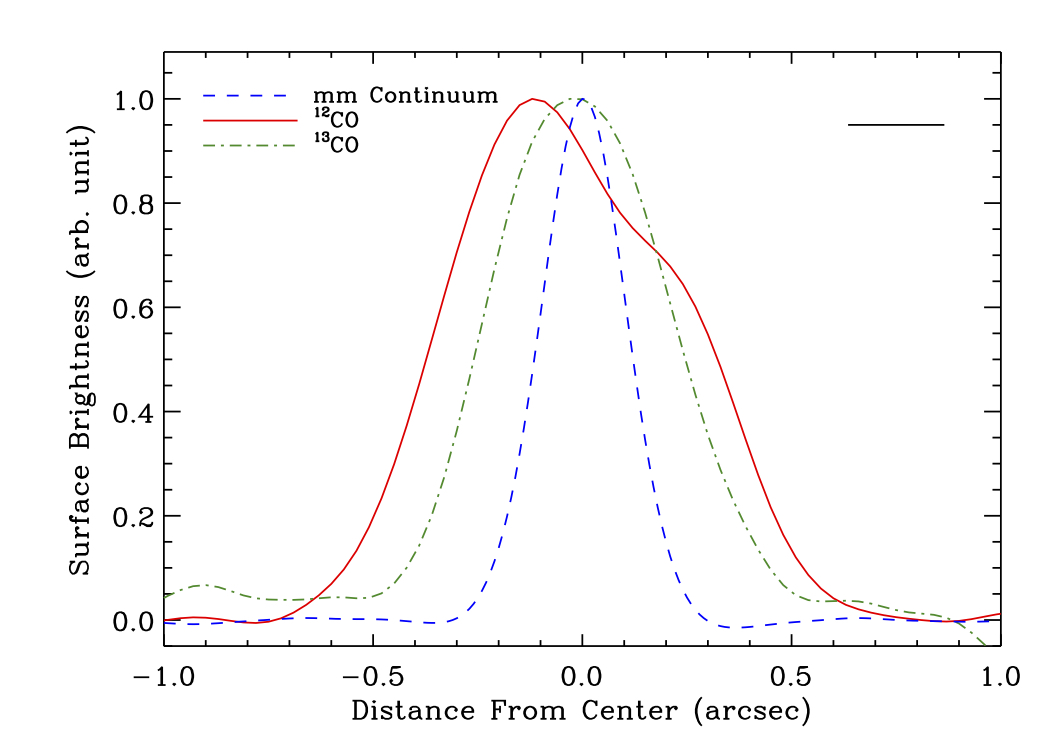}
\caption{Brightness profiles along the major (left panel) and minor axis (right panel), integrated over the other dimension in both cases, of the dust millimeter continuum, $^{12}$CO, and $^{13}$CO data of Oph163131 (dashed blue, solid red and dot-dashed green, respectively). The horizontal $^{13}$CO profile has been convolved by a Gaussian with a width equal to the minor axis of the beam to improve signal-to-noise.
All brightness profiles are displayed in arbitrary units, after normalization to their respective maxima. The horizontal bar in the top right represents the FWHM of the beam along its major axis.}
\label{fig:ALMA_cont_vs_gas_cuts}
\end{figure*}

In Figure \ref{fig:ALMA_cont_vs_gas_cuts}, we present a comparison between $^{12}$CO, $^{13}$CO, and 1.3 mm dust continuum brightness profiles along the minor and major axes of the disk. The left panel of Figure \ref{fig:ALMA_cont_vs_gas_cuts} shows that the dust disk is smaller radially by a factor of almost 2 than the  $^{12}$CO gas. Similar results where the gas component extends further in radial direction than the dust component have been reported in the literature and are interpreted as inward radial drift of dust grains \citep{Ansdell2018}. We note that the double peak component in the integrated gas profile is the natural result of the high inclination of the disk, with some self-absorption occurring along the line of sight to the central star. \cite{Villenave2020} measured a full radial extent of $2\farcs50 \pm 0\farcs01$ for the disk based on the continuum map (see also Paper\,I). The right panel of Figure \ref{fig:ALMA_cont_vs_gas_cuts} demonstrates that the dusty disk is vertically thinner than both gas components, and possibly much thinner as the continuum is unresolved in this direction. From this panel, one can also see an asymmetric profile in the $^{12}$CO, caused by the brightness differences between the SE and NW layers, but a much more symmetric profile for the $^{13}$CO component, supporting the fact that no significant structure is detected in $^{13}$CO.

\section{Tomographic Gas Temperature Retrieval} 
\label{sec:model}
\subsection{Method}
Temperature estimates of protoplanetary disks are often obtained through detailed modeling of (sub)-millimeter observations or from long wavelength coverage SED fitting \citep[e.g.,][]{Woitke2019}. In both of these approaches, it is common to parameterize the temperature and the density structure of a disk and then vary many of these parameters until a good match between models and observations is reached. These many parameter fitting techniques are often under-constrained given the real complexity found in high-angular resolution disk observations.

Direct measurement of the temperature structure of disks is therefore not only important in our understanding of the radiative processes of disks, but it can, in principle, also be used to constrain  current modeling work.  Here we describe a model independent technique developed to measure the temperature structure of protoplanetary disks. This technique, referred to as Tomographically Reconstructed Distribution (TRD) was first implemented by \cite{Dutrey2017} and was applied to the edge-on protoplanetary disk `The Flying Saucer'. A similar method was recently implemented by \cite{Teague2020} for Gomez' Hamburger (2MASS J18091339-3210500).

The TRD technique is a geometrical approach designed to identify parcels of gas on a spatially and spectrally resolved protoplanetary disk and map these parcels of gas onto a 2D radial and vertical structure. This technique assumes that the disk's inclination angle is close to edge-on and that the disk follows a Keplerian velocity field. Although the latter point is not exactly correct, as gaseous disks are pressure supported systems, and therefore they do not rotate exactly at Keplerian velocities. It has been demonstrated that omitting the pressure term ($-\frac{1}{\rho}\frac{dP}{dr}$) only causes a small over-estimation of the gas velocity at large disk radii. \cite{Pinte2018} analyzed the velocity pattern of the disk around IM~Lupi (whose mass is similar to that of Oph163131) and found that the pressure gradient term only increases the rotational velocity of the disk by 0.1 $\rm km \, s^{-1}$ at distances larger than 300 au. Since our disk does not extend beyond 300~au and our velocity resolution is already larger than this effect, we are safe by assuming pure Keplerian rotation.

In order to convert the incident flux emitted from the gaseous component of the disk into brightness temperature, we assume that the gas is in local thermodynamic equilibrium (LTE). The LTE condition is valid in high density regions, for example, towards the midplane of most protoplanetary disks, but not necessarily towards the disk's photosphere. Besides, when the emission is optically thick and fills the beam, the brightness temperature becomes comparable to the gas's excitation temperature. This is typically the case for $^{12}$CO, but it can also be true for the rarer isotopologue $^{13}$CO \citep[see for example ][]{Pinte2018}. Furthermore, the excitation temperature of low-J CO lines, as is our case here, follows very closely the kinematic temperature of the gas \citep{Pavlyuchenkov2007}.

In the following paragraphs, we briefly describe the steps involved in the reconstruction of the temperature distribution of a protoplanetary disk. We refer the reader to Appendix \ref{sec:Equations} for the derivation of the equations we used to reconstruct the disk emission.

Our modeling technique depends on the stellar mass ($\rm M_{\ast}$), the distance to the source ($d$), and the inclination angle of the disk relative to the observer ($i$). These values are input into equations (\ref{eq:rho_equation}) and (\ref{eq:z_equation}), which parameterize the radial position ($\rho$), and the altitude above the mid-plane ($z$) of small parcels of gas in a protoplanetary disk as a function of the observed line of sight velocity ($v_{y}$), the impact parameter ($x$), and the projected altitude above the mid-plane ($z'$). A PV diagram is then selected at a projected altitude $z'$, and the code calculates the radial $\rho$ and altitude $z$ coordinates, in addition to the intensity value $I$ of the disk at such locations.  This process is repeated by selecting a new PV cut at a different projected elevation ($z' + \Delta z'$) to obtain a new set of intensity values for all the radial positions $\rho$ and elevations ($z + \Delta z$). This iteration continues until all the projected altitudes above and below the projected midplane are analyzed.

\begin{figure*}[ht!]
\epsscale{1.15}
\plotone{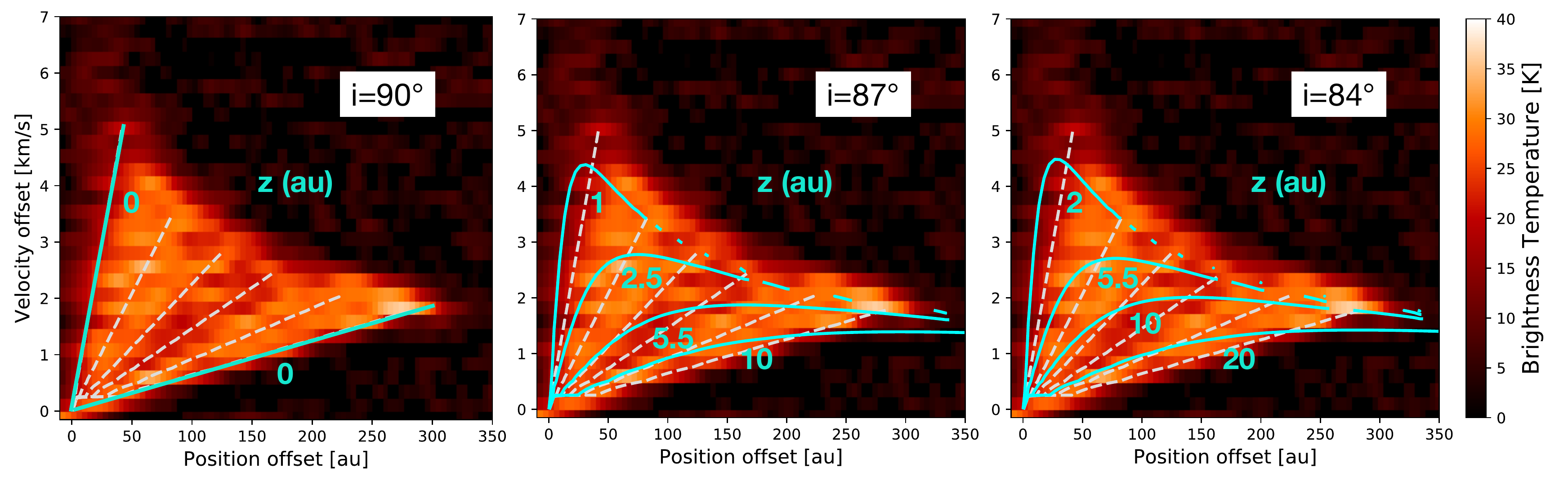}
\caption{Demonstration of the TRD method assuming three different disk's inclination angles: $i=90\degr$ (left), $i=87\degr$ (middle), and $i=84\degr$ (right). All three figures show the same PV diagram along the disk's major axis. White dashed lines correspond to iso-radial lines ranging from 40 to 300 au, these lines remain the same with inclination. Cyan solid lines represent iso-altitude lines (regions at the same vertical distance above the midplane). Assuming a perfectly edge-on disk (left panel) the flux from different radii emerge at a fixed altitude ($z=0$~au in this case). As the disk is declined out of the edge-on position, the observed flux displays a radial and vertical dependency within the PV cut. At an inclination of $i=87\degr$ (middle panel) the emission emerges from altitudes ranging between $1 \lesssim z \lesssim 10$ au while at an inclination of $i=84\degr$ (right panel) the emission arises from even higher altitudes, $2 \lesssim z \lesssim 20$ au.}
\label{fig:model-explanation}
\end{figure*}

It is worth noting that in a perfectly edge-on case ($i=90\degr$), the emission obtained from a PV diagram emerges from a fixed elevation in the disk, allowing us to construct a one-to-one mapping between projected and true mid-plane positions. In cases where $i \neq 90\degr$, however, the emission from a PV diagram obtained at a given projected elevation $z'$, comes from a range of elevations $\Delta z$ above and below such projected elevation. This is exemplified in Figure \ref{fig:model-explanation} where we show the same PV diagram, measured at $z' = 0$, but we use three different disk inclination angles. In the left panel, where $i=90\degr$, the emission from different radial positions are expected to arise from a unique vertical elevation $z=0$~au. When the inclination is assumed to be $i=87\degr$ (middle panel) the emission from different radial positions emerge from a variety of elevations above the midplane $1 \lesssim z \lesssim 10$ au. Lastly, if the inclination is assumed to be $i=84\degr$, then the range of elevations where the emission is assumed to be emitted, is even larger than before $2 \lesssim z \lesssim 20$ au. Therefore, we emphasize the importance of considering small deviations out of edge-on inclinations when using the TRD technique.

After all the intensity values $I$ and spatial coordinates $\rho$ and $z$ from the PV cuts are collected, we construct a two dimensional array of radial (\textit{R}) and vertical (\textit{Z}) positions with steps of 10 au in both directions. This spacing size was chosen to oversample by factors of $\sim3$ and $\sim4$ the resolution of our observations in the radial and vertical directions, respectively. Intensity values are then assigned to each of these \textit{R-Z} cells and median values are calculated within each cell. 
To calculate the statistical uncertainty in our method, we additionally measure the 16th and 84th percentiles of the intensity values for each of the cells, and find a $2-4$~mJy ($1-2$~K temperature) uncertainty in most parts of the disk. Before the median intensity is converted into brightness temperature, each of the \textit{R-Z} cells with 10 or fewer individual intensity values or cells with median intensity below 9 mJy (or 4.5~K) were masked. This ensures that the statistical computations of the median, the standard deviation, and the percentiles are meaningful. At last, the average intensity values are transformed into brightness temperature using Planck's equation (\ref{eq:brightness_temp}), where T$_{bg}$ is the microwave background emission of 2.73 K.

\begin{equation}
\label{eq:brightness_temp}
    \rm T_b = \frac{h \nu}{k} \left[ \ln{ \left( \frac{2 h \nu^3}{I_{\nu}c^2}+1\right)} \right] ^{-1} + T_{bg}
\end{equation}

In Appendix \ref{sec:model_testing}, we show a variety of tests performed using synthetic disk models computed with MCFOST. These tests, which include disk's inclination and spatial resolution  variations, confirm that the temperature retrieval method works well in the interior regions of the disk where the density is large enough to meet our optical depth criteria.


\subsection{The temperature structure of Oph163131}
\label{subsec:temperature_of_oph163131}

\begin{figure*}[ht!]
\plotone{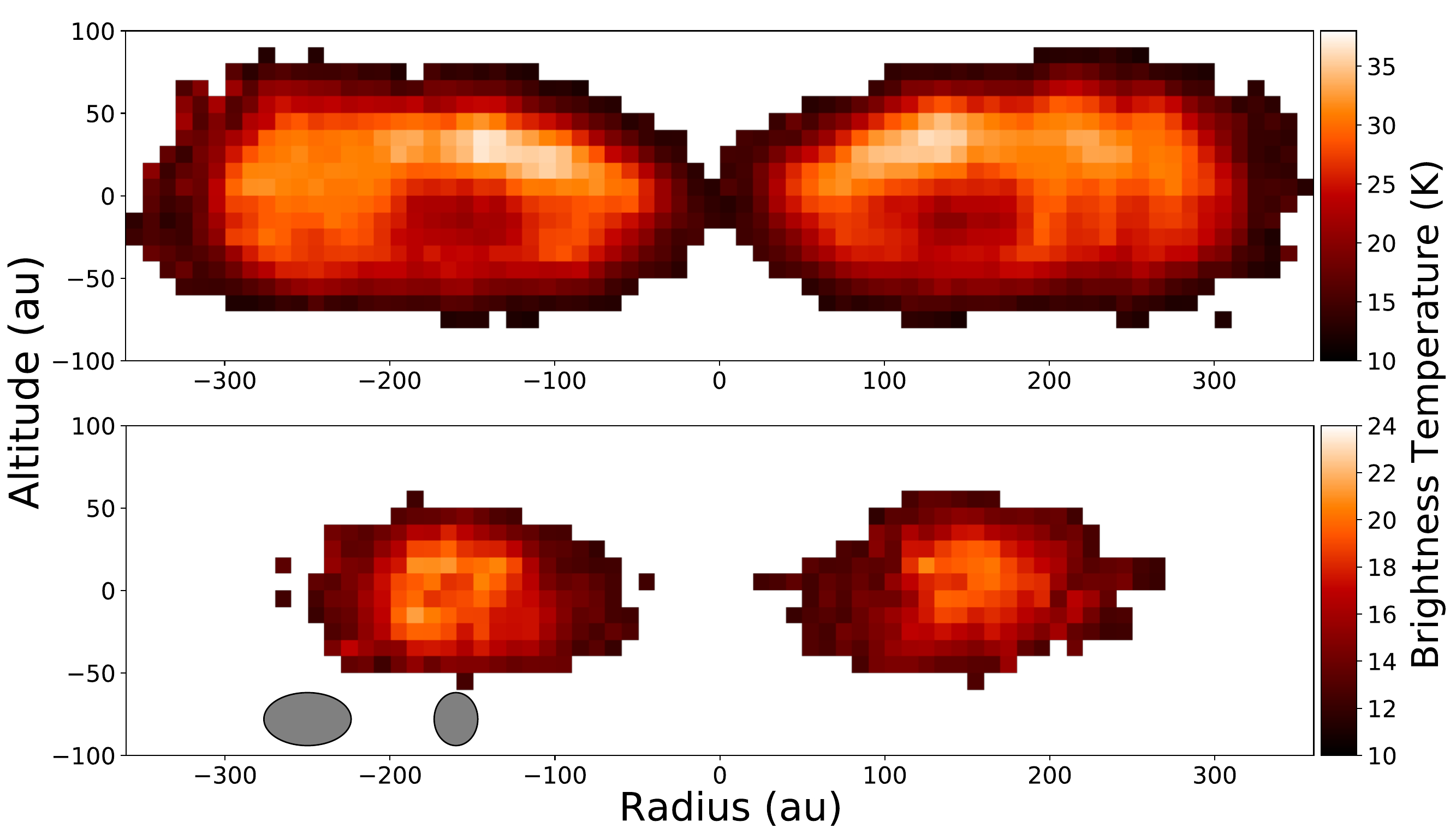}
\caption{Tomographically reconstructed distribution of the $\rm ^{12}CO$ (top panel) and the $\rm ^{13}CO$ (bottom panel) lines. The reconstructed temperature maps are shown in physical units of radius and altitude, both in au. On the $\rm ^{12}CO$ reconstructed temperature, we distinguish the near and the far emitting layers of the disk. The $\rm ^{13}CO$ reconstructed temperature, on the other hand, does not show any significant structure. The grey ellipses shown on the top left part of the bottom panel illustrate how the Keplerian velocity shear affect the resolution as a function of radius.}
\label{fig:TRD_12CO}
\end{figure*}

We now apply the TRD method to the $^{12}$CO~J~=~2~-~1 and $^{13}$CO~J~=~2~-~1 data cubes of Oph163131\footnote{We also applied this method to the non-continuum subtracted data cubes, but no significant differences were found.}. We fixed the stellar mass to 1.2 $\rm M_\odot$, the disk inclination angle to $i=87\degr$, and the distance to the source to $d=147$~pc. The top panel of Figure \ref{fig:TRD_12CO} shows the result of the reconstructed temperature distribution for the $^{12}$CO line. The reconstruction technique was applied to both sides of the disk (positive and negative velocities with respect to $v_{sys}$), both of which show a maximum radial extent of $\rm R_{max} = 350$~au and altitude above the midplane of $\rm Z_{max} = 80$~au. From this plot, we can clearly recognize the `top' and `bottom' layers of the disk, which are separated by a colder midplane that extends from about 100 au to 200 au in radius. The peak brightness temperature on both sides of the disk reaches about the same temperature of $35\pm 2$~K at a distance of 135 $\pm$ 20~au from the center of the disk. The displacement of the peak brightness temperature outside of the inner part of the disk is mostly caused by the beam convolution effects which are described in Appendix \ref{sec:model_testing}. However, towards the innermost regions of the disk $\rm R\lesssim 1 $~au (which cannot be accessed at our spatial resolution) dust absorption could also play a role in decreasing the apparent brightness temperature of the gas and therefore shift the peak temperature position. 

In the disk, emission detected at radial distances smaller than $\sim130$~au are more heavily affected by beam dilution due to the smaller scale height at the inner parts of a flared disk. As demonstrated in Appendix \ref{subsec:models_different_resolutions}, beam convolution results in underestimating the maximum temperature of the disk and overestimating the gas temperature close to the midplane. The temperature gradient from the $^{12}$CO reconstructed temperature is therefore only a lower limit to the true temperature gradient in Oph163131.

The bottom panel of Figure \ref{fig:TRD_12CO} shows the reconstructed temperature distribution for the $^{13}$CO isotopologue. This reconstructed distribution is radially and vertically more compact than the $^{12}$CO temperature distribution, and we only detect emission at vertical distances smaller than 45 au and radial distances between 60 au and 250 au. A similar reduction in the relative height of the CO isotopologue emission layers was identified by \cite{Teague2020} for Gomez' Hamburger. A combination of lower S/N, lower optical depth, and beam smearing prevent us from identifying clear structures in this isotopologue. The detected $^{13}$CO emission is unlikely to arise from the midplane due to CO freeze-out; instead, the emission we detect is likely emitted by thin layers just above and below the freeze-out zone, which are completely unresolved by our limited $\sim0$\farcs 2 spatial resolution. 

On the lower left part of Figure \ref{fig:TRD_12CO}, two gray ellipses represent the spatial resolution variations at different radial distances. These radial resolution differences are due to the Keplerian shear velocity of the disk, calculated as $dr = 2r\,dv/v(r)$, where $v(r)$ is the Keplerian velocity of the disk and $dv$ is the local line width. For distances smaller than 160 au, the beam size dominates the radial resolution limit of the map, however, at distances larger than 160 au the radial resolution is limited by the Keplerian velocity shear. The two ellipses shown at 160 au and 250 au, represent then radial resolution variations from 30 au and 60 au. We note that the local line width in our dataset is dominated by our velocity resolution of 0.25 $\rm km \, s^{-1}$, and not by the combined thermal and turbulent line broadening, as was the case for the analysis performed on the Flying Saucer \citep{Dutrey2017}.

\subsection{Radial and vertical temperature profiles}
\label{subsec:radial_and_vertical_profiles}

\begin{figure*}[ht!]
\epsscale{1.15}
\plotone{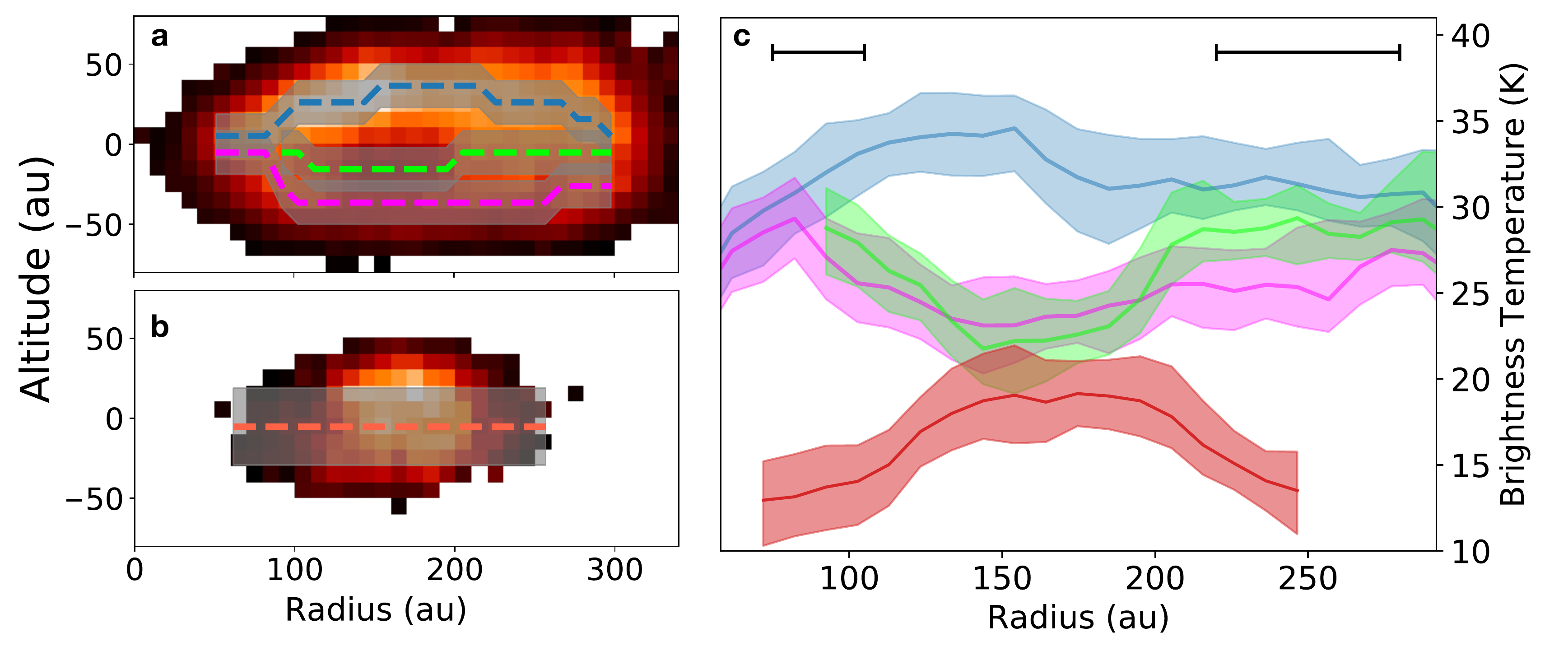}
\caption{Panels a and b show the left-right averaged temperature distribution from Figure \ref{fig:TRD_12CO}. Color lines correspond to a visual trace over the near, far and cold midplane layers for the $\rm ^{12}CO$, and just center for the $\rm ^{13}CO$. The wider transparent region next to the white traces correspond to the three pixel (five pixel) vertical average used to create panel c for $\rm ^{12}CO$ and $\rm ^{13}CO$, respectively. Panel c shows the traced temperatures as a function of radius for the cold midplane, bottom and top layers of the $\rm ^{12}CO$ map, and the center temperature for the $\rm ^{13}CO$ map. The shaded regions around the solid lines, correspond to the temperature uncertainties for each layer. Horizontal bars in panel c, drawn at distances of 100 au and 250 au, depict the changing radial resolution of this method (see subsection \ref{subsec:temperature_of_oph163131})}
\label{fig:temperature_vs_radii}
\end{figure*}

To better understand the radial temperature structure of the disk, we show the temperature variation as a function of radius for $^{12}$CO and $^{13}$CO in Figure \ref{fig:temperature_vs_radii}. We first averaged both sides (left and right in Figure \ref{fig:TRD_12CO}) of the temperatures maps, and then visually identified three distinct layers for the $^{12}$CO (panel a) and a single layer going through the center of the $^{13}$CO (panel b). A three pixels average in the vertical direction (24 au) for the $^{12}$CO, and a five pixels average (40 au) for the $^{13}$CO were used to compute the temperature profiles shown in panel~c. 

Panel c of Figure \ref{fig:temperature_vs_radii}, shows that the temperature of the top  layer (in blue) increases from $27$~K at 60~au to a peak temperature of $35$~K at 150~au, then it declines to a temperature of roughly $31$~K at a distance of 210~au and remains constant towards the outer part of the disk. The initial increase in temperature as a function of radius is not physical, but mostly caused by beam dilution (as discussed in Section \ref{subsec:temperature_of_oph163131}). On the other hand, the subsequent temperature decline is meaningful and allows us to understand how rapidly the temperature change as a function of radius. The bottom (purple) and cold (green) layers share very similar temperature profiles. The bottom layer reaches a minimum temperature of  $23$~K at 150~au and it increases toward the outer edge of the disk to a steady temperature of  $25$~K beyond 210~au. The cold midplane, on the other hand, reaches a minimum temperature of $21$~K at 150~au and then increases to a temperature of 28~K beyond 210~au. The temperature increase of the cold midplane region towards the outer parts of the disk is an interesting feature and is further discussed in subsection \ref{subsec:the_outer_disk}.

The $^{13}$CO temperature profile (in red) reaches a maximum temperature of $19$~K at distances of 150 to 200 au, and due to beam dilution, it decreases toward shorter and longer radial distances. The lower temperature found for $^{13}$CO compared to $^{12}$CO likely results from the combination of 1) a lower optical depth (similar to, or lower than, unity), 2) a smaller vertical extent, whereby the two emitting layers of $^{13}$CO (above and below the presumed frozen-out midplane) are only partially resolved by our observations, and 3) a lower temperature in these more embedded layers. 


\begin{figure}[ht]
\plotone{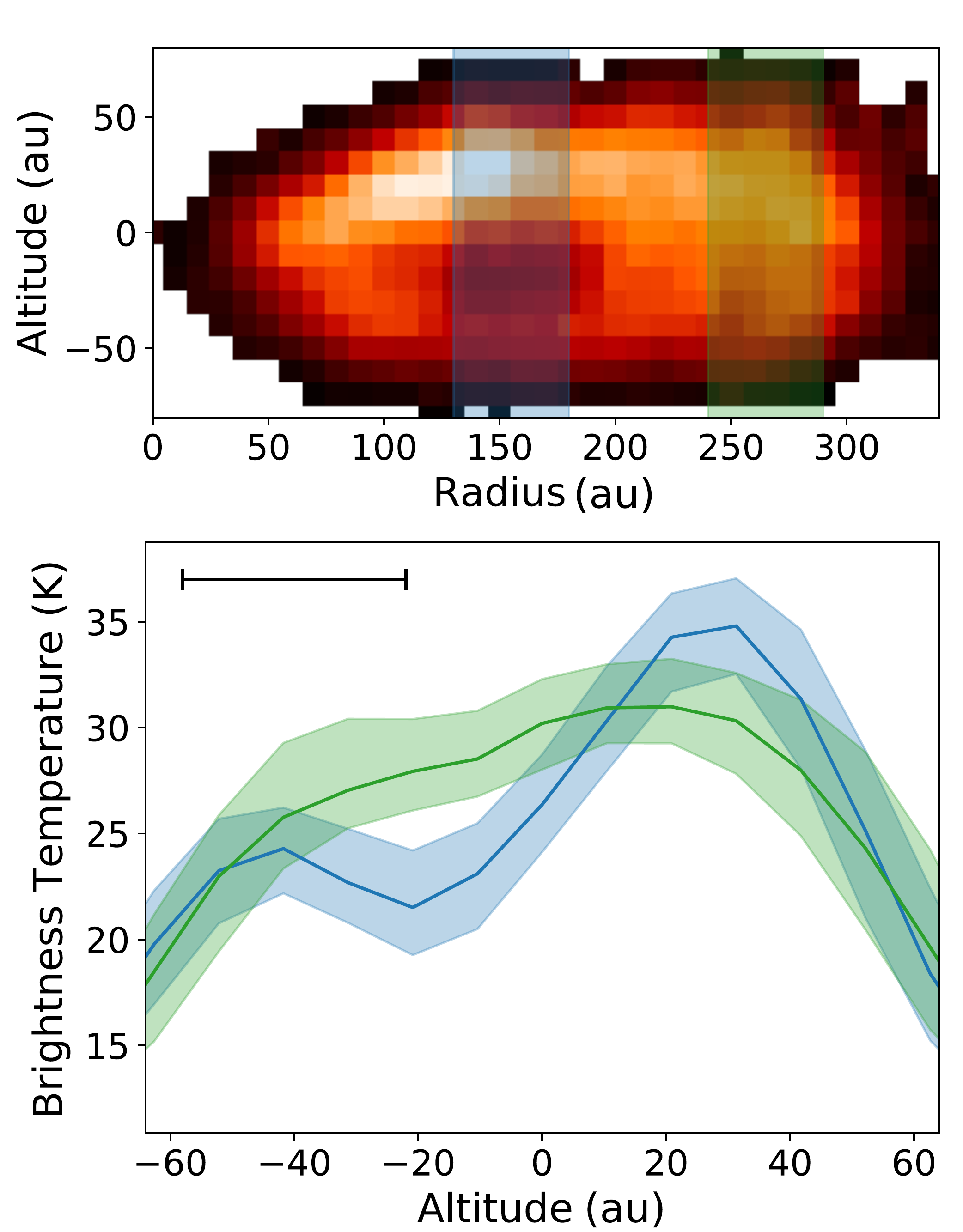}
\caption{Top panel: left-right average of the $^{12}$CO temperature map. The two color shaded regions at distances of 135-175 (blue) and 245-285 (green) mark the distances where radial averages are performed. Bottom panel: temperature as a function of the disk's altitude taken at 135-175 (blue) and 245-285 (green) as shown in the top panel. The shaded regions around the solid lines, correspond to the temperature uncertainties calculated from the maps. The horizontal bar shown in the bottom panel corresponds to the spatial resolution of 36 au in this direction.}
\label{fig:temperature_vs_vertical}
\end{figure}

Although it seems clear from Figure \ref{fig:TRD_12CO} that in $^{12}$CO  there are a top and a bottom emitting layers separated by a colder midplane, the three pixel averaged profiles of Figure \ref{fig:temperature_vs_radii} suggest that the midplane is not significantly cooler than the bottom layer. We further investigate this ambiguity by looking at vertical temperature profiles of the $^{12}$CO map. As shown in  Figure \ref{fig:temperature_vs_vertical}, we chose two regions to perform  the vertical cuts, one region at a radial distance of $135-175$ au, and a second region at radial a distance of $245-285$ au. The first region was chosen to go through the cold midplane region in the disk (marked as a light blue region) while the second one was chosen to go through the outer disk (marked as a light green region). In the bottom panel of Figure \ref{fig:temperature_vs_vertical}, we show the temperature of the disk as a function of vertical distance above and below the midplane in the two regions marked above. The blue line clearly depicts the top and bottom emitting regions and the colder midplane zone mentioned before, with the midplane region being $\delta T=3 \pm 2$ K cooler than the bottom region and $13\pm2$ K cooler than the top region. This is qualitatively consistent with TRD maps of the Flying Saucer and Gomez' Hamburger \citep{Dutrey2017, Teague2020}. The temperature structure changes significantly when the cut is performed towards the outer region. The green curve does not show a valley structure, which shows that the midplane temperature has increased at radial distances larger than $R\sim200$ au, but there is still a marginal level of asymmetry showing that the top side of the disk still appears to be slightly warmer than the bottom side at larger distances from the center. 

The temperature asymmetry between the top and bottom sides of the disk is an optical depth effect caused by a slight deviation out of edge-on inclination of Oph163131 coupled with the significant line optical depth and vertical thickness of the warm molecular layer. At the top of the disk ($z>0$), the optical depth $\tau=1$ is reached towards the disk's surface, therefore providing the temperature of the disk's high elevation layers which is warmer as it is directly exposed to the star. This effect is further amplified by the fact that, even at small deviations out of edge-on position, the far side of the disk, on the top side, becomes visible to the observer (even if the disk is not resolved), causing a larger area of the disk to be exposed to the observer and consequently artificially increasing the measured temperature (see Appendix \ref{subsec:models_different_inclinations}). On the bottom side ($z<0$ au) of the disk, however, the $\tau=1$ surface is reached closer to the midplane from the observer's perspective, providing a temperature measurement of the disk's lower elevation regions. Additionally, it is expected that at the locations where the dust continuum is optically thick, the gas emission could be absorbed, further lowering the brightness temperature. Given the geometry of the disk, this effect only plays a role in regions below the midplane ($z<0$). This is why even after the TRD technique transforms the disk into $R-Z$ coordinates, there is still a temperature asymmetry between the two sides of the disk.

The temperature asymmetry, combined with the angular resolution of the observations and the intrinsic angular scales of the Oph163131 disk, also explains the off-centered cold disk midplane (by approximately $-20$~au) shown in Figure \ref{fig:temperature_vs_vertical}. The more exposed CO upper emitting layer appears broader and hotter than the less-exposed CO bottom emitting layer, which causes the whole disk midplane to be shifted towards the bottom side of the disk (see also Figure \ref{fig:Changes_in_inclination} of the Appendix). All of the above is then amplified by the flux smearing induced by the spatial resolution of the observations. Finally, the close match between the temperature of the bottom $^{12}$CO emission and of the (symmetric) $^{13}$CO is consistent with observations of IM\,Lup \citep{Pinte2018} and confirm a freeze-out temperature of $\sim$20\,K.

\section{Discussion}
\label{sec:discussion}
\subsection{Oph163131 compared to other sources}
We now put Oph163131 in context by comparing its properties against other sources with direct disk temperature measurements. We focus on four sources with well spectrally and spatially resolved $^{12}$CO observations, where a model-independent temperature measurement was performed: IM Lupi \citep{Pinte2018}, the Flying Saucer \citep{Dutrey2017}, Gomez's Hamburger \citep{Teague2020}, and HD~163296 \citep{Dullemond2020, Isella2018}. Two of the sources (Gomez's Hamburger and the Flying Saucer) harbor edge-on disks and have been analyzed using the TRD method in a very similar fashion as we have done here for Oph163131, therefore a full reconstructed temperature map is available for comparison. The other two sources (IM Lupi and HD~163296) have disks with intermediate inclination angles ($i\sim45\degr$) and report direct temperature measurements on specific parts of the disk (i.e., at the surface level or towards the midplane).

The central stars of these sources span a relatively large range of masses, Gomez's Hamburger and HD~163296 are regarded as Herbig stars with stellar masses of $2.5\pm 0.5$~$\rm M_\odot$ \citep{Teague2020} and $2.04^{+0.25}_{-0.14}$~$\rm M_\odot$ \citep{Andrews2018}, respectively. IM Lupi and the Flying Saucer, on the other hand, have stellar masses of $1.0\pm0.1$~$\rm M_\odot$ and $0.58\pm0.01$~$\rm M_\odot$, respectively, both consistent with being T~Tauri stars, which make them better candidates to be compared against the disk of Oph163131. The much higher luminosity of Herbig stars should result in substantially higher temperatures throughout the disk \citep[e.g.,][]{Kamp2011}. 

The maximum temperature measured from $^{12}$CO observations in the  disks of the  Herbig star's Gomez's Hamburger and HD~163296, reach values of $\sim 60$~K and $\sim90$~K, respectively, which are significantly hotter than the temperatures measured in the T~Tauri stars' disks of $\sim30$~K for IM Lupi and Oph163131, and of $\sim20$~K in the disk of the less massive star Flying Saucer. Minimum temperatures of the disks are harder to compare as they strongly depend on where the temperatures are measured, how well resolved the disks are, and the sensitivity of the observations. 

\cite{Pinte2018} produced radial temperature profiles computed from the $^{12}$CO and the $^{13}$CO lines for IM~Lupi (their figure 4, right panel), which can be easily compared to our Figure \ref{fig:temperature_vs_radii}. In $^{12}$CO, the maximum temperatures of the disks are 35~K and 32~K which are reached at distances of $\sim 150$~au and $\sim 190$~au for Oph163131 and IM~Lupi, respectively. The  temperature decline towards larger radii, however, is markedly shallower in Oph163131 than in IM~Lupi: $\sim2.5$~K per 100~au and $\sim6$~K per 100~au, respectively. In $^{13}$CO, the radial profiles of both sources reach a plateau with peak values of 19~K and 21~K for Oph163131 and IM~Lupi, with similar decreases towards the outermost regions of the disks. We note, however, that contrary to the IM~Lupi study, our $^{13}$CO observations do not fully resolve the layers on either side of the midplane and, thus 
the $^{13}$CO temperature profile reported in our study is only a vertical average, making the comparison less straightforward.

For most of the sources mentioned above, we have estimated the ratio between the disk's full vertical extent ($Z_{max}$) and total disk's  radial size ($R_{max}$) in $^{12}$CO. The thicker disk, using this metric, is the Flying Saucer with a ratio of $\frac{Z_{max}}{R_{max}}\sim0.5$, IM~Lupi and the Gomez's Hamburger come next with a similar vertical to radial ratio of $\frac{Z_{max}}{R_{max}}\sim0.33$, and finally the thinnest disk is Oph163131 with a ratio of $\frac{Z_{max}}{R_{max}}\sim0.23$. In this comparison, we have excluded HD~163296, as the vertical size measurement performed on this disk is strongly model dependent and therefore not a fair comparison against the other sources \citep{Isella2018}. 

Despite contrasting disks sizes of sources obtained at different spatial resolutions,  these comparisons can provide some insights into how the stellar masses, evolution, and the temperature of the disks relate to the aspect ratio of the sources. For T~Tauri stars, it appears that more massive stars correlate with thinner disks, although with only three sources in the sample, this result lacks statistical significance. An alternative view is that each disk could be in a different evolutionary state where vertical settling may have progressed by different amounts. In the case of Oph163131, the absence of strong accretion (see Section \ref{subsec:sepctral_characterization}) means that there is much less UV radiation available to heat up the gas, which ultimately causes the flat aspect of this disk. \cite{Alcala2017A} measured an accretion luminosity of 0.08\,$L_\odot$ for IM Lupi by modeling the excess emission from the UV to the near-infrared of X-shooter data. The excess of accretion radiation might explain the larger $\frac{Z_{max}}{R_{max}}$ value for IM Lupi when compared to Oph163131. As for the Flying Saucer, we have not found accretion measurements in the literature. HST images show no signs of any jet or molecular outflows in the vicinity of this source, yet this does not rule out stellar accretion. 

The accretion status of other edge-on protoplanetary disks that have a low height to radius ratio and display flat parallel nebulae in scattered light images (as seen in Oph1613131) is discussed next. HK\,Tau\,B, a well studied edge-on disk \citep{Koresko1998, Stapelfeldt1998}, was classified as a CTTS by \cite{Monin1998}, albeit with an optical spectrum that displays a modest $H_\alpha$ emission and no other emission line.
The optical spectrum of LkH$\alpha$\,263C, a similarly flat disk \citep{Jayawardhana2002, Chauvin2002}, 
reveals that it has a strong $H_\alpha$ emission along with prominent forbidden emission lines. Finally ESO\,H$\alpha$\,574, another flat-shaped disk imaged with HST \citep{Stapelfeldt2014}, has a prominent optical jet, which suggests ongoing stellar accretion. In summary, flat shaped proplanetary disks are found around actively accreting and non-substantially accreting stars. We encourage a study that would quantify the accretion radiation and aspect ratio of disks to test whether accretion-induced UV radiation could significantly affect the vertical thickness of protoplanetary disks.

\subsection{The inner disk: inside 200 au}
\label{susbsec:the_inner_disk}

\begin{figure*}[ht!]
\plotone{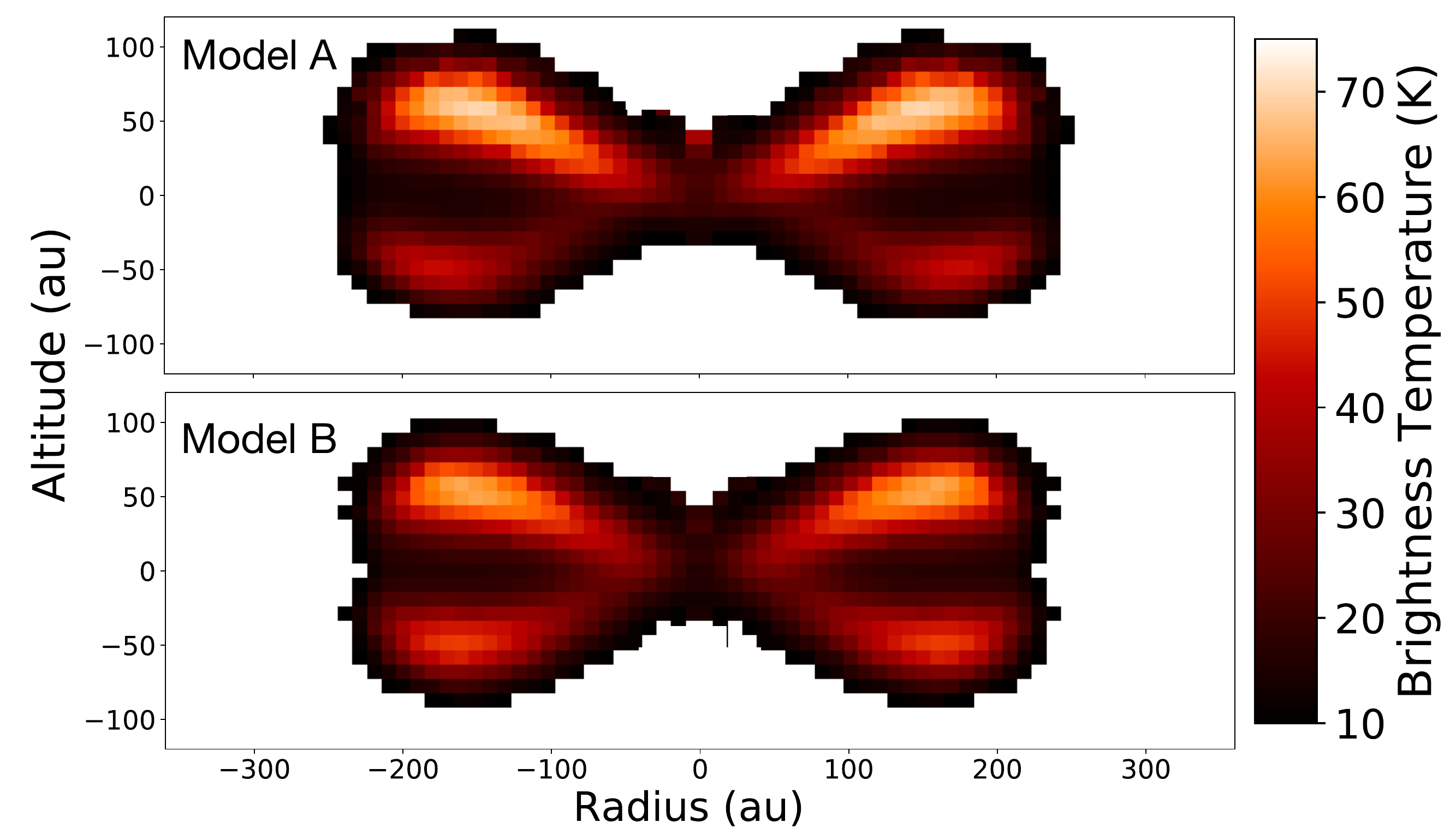}
\caption{Comparison between the two temperature maps produced using the MCFOST synthetic models with parameters adopted from Paper\,I. The top panel corresponds to the temperature map obtained using the 0.8 \micron \, HST best fit parameters (Model A), and the bottom panel from ALMA 1.3 mm continuum parameters (Model B).}
\label{fig:Models_vs_data}
\end{figure*}


The reconstructed $^{12}$CO temperature map of Oph163131 shows two distinct regions with a demarcation at a radius of $\sim200$\,au. Inside of this radius, the CO emission arises from two elevated layers separated by a midplane, whereas the outer region consists of a much more vertically uniform emission. Interestingly, this radial break coincides with the outer radius at which both the submillimeter continuum emission and the scattered light are detected (see Figure \ref{fig:HST_ALMA_cont} and also \cite{Villenave2020}. We, therefore, address the ``inner" and ``outer" regions separately. Considering first the inner disk, we note that the observed behavior of the gas emission is similar to that observed in other disks (e.g., IM~Lupi and The~Flying~Saucer discussed above). We now wish to go beyond qualitative comparisons and evaluate whether detailed radiative transfer models of the dust components of Oph163131 can reproduce the temperature structure we obtain from the gas.

In Paper\,I, ALMA 1.3 mm dust continuum observations and 0.8 \micron \, HST scattered light images of Oph163131 were modeled. To reproduce the observations, Paper\, I uses the MCFOST radiative transfer code and a Markov Chain Monte Carlo (MCMC) approach to fit the millimeter and sub-micron data of Oph163131 independently. The individual fits, which assumed a tapered-edge surface density profile, provided a set of density and geometrical parameters of the disk such as inclination angle, scale height, surface density, flaring exponent, among others (see their Equation 1 and Table 4). In this section, we compare the models from Paper\,I to our $^{12}$CO reconstructed temperature map, limiting the analysis to the inner 200\,au. 

To perform a meaningful comparison, we took the following approach. We used Paper I dust disk density and geometrical parameters, a gas-to-dust ratio of 100, a CO/H$_2$ abundance of $10^{-4}$, and we imposed CO freeze out (CO/H$_2$ = 0) in regions where the temperature of the disk falls below 20~K. We then computed the synthetic $^{12}$CO~J~=~2~-~1 observations of such a disk with MCFOST. From the resulting data cubes, we analyze these models in the exact same manner as we did with gas observations in Section \ref{sec:model}, i.e., using the same parameters for Oph163131 such as stellar mass, spatial and spectral resolution, distance to the star, etc. The two models we consider are a) a model that fits the HST 0.814 \micron \, image (Paper\,I's model A) which has a disk inclination angle of $i=84.5\degr$, a scale height of H$_0$~$=7.2$~au (at 100 au), and a critical radius of $R_{C}=89$~au, where $R_{C}$ has the usual meaning of the distance at which the disk surface density changes from a power-law to an exponential taper; and b) a model that reproduces the 1.3 millimeter continuum ALMA map (Paper\,I's model B), with an inclination angle of $i=87.2\degr$, a scale height of H$_0$~$=9.5$~au (at 100 au), and a critical radius of $R_{C}=74.5$~au. The model retrieved temperature maps are shown in Figure \ref{fig:Models_vs_data}.

Both models produce a disk with clearly defined top and bottom CO emission layers separated by cold and frozen midplane. The maximum temperature of the gas, however, is almost twice as hot as the data, and the elevation of the CO layers in the 150 to 200~au radial range is also significantly higher (50~au) than the observations ($\sim35$~au). This means that the parameters that reproduce sub-micron (Model A) and millimeter size (Model B) dust grains do not necessarily match the gas observations when adopting standard gas-to-dust and CO-to-$\rm H_2$ abundance ratios.

One possible path to reconcile the dust and gas observations is to reduce the overall CO gas mass in the disk. This would decrease the altitude at which the CO becomes optically thick and also the temperature of the emitting layer, as the gas is found at deeper layers in the disk. In practice, this means to either adopt a lower gas-to-dust ratio ($<100$) or a lower CO-to-$\rm H_2$ abundance ratio ($<10^{-4}$) in the disk. \cite{Ansdell2016} performed a large study of young stars in Lupus and found that most disks have gas-to-dust ratios of less than 100 for an assumed standard CO-to-$\rm H_2$ abundance ($10^{-4}$). \cite{Long2017} analyzed a large number of disks bearing sources in Chamaleon I and found median gas-to-dust ratios of $\sim4 - 15$, depending on the assumptions made. Recent observations of individual disks have shown that CO could be depleted by several orders of magnitude with respect to H$_2$ \citep{Favre2013,Kama2016,Trapman2017,Zhang2017,Zhang2020}. These anomalous abundances have been explored using physical and chemical models that deplete CO in disks \citep[e.g.,][]{Bosman2018}. In particular,  \cite{Krijt2020} used models that combine chemical processing of CO with ice sequestration in the disk midplane to obtain strong CO depletion in the outer disk warm molecular layers.

Although there is extensive literature that suggests the possibility of a lower gas-to-dust ratio and CO-to-$\rm H_2$ abundance in disks, such ad-hoc adjustment in our models of Oph163131 does not necessarily provide a satisfactory match to the gas observations. One example of this is that using a lower gas-to-dust ratio would immediately affect the dust settling mechanism, and just like this, other backreactions could occur if lower CO-to-$\rm H_2$ abundances are adopted. As discussed in the next subsection, only detailed modeling of the CO observations would provide the appropriate physical parameters of the gas component of the disk around Oph163131.

\subsection{The outer disk: outside 200 au}
\label{subsec:the_outer_disk}
In subsection \ref{subsec:radial_and_vertical_profiles}, we pointed out that the outer disk ($\rm R >200$~au) of Oph163131 has an almost constant temperature of $\sim$30 K at all elevations and radii. This is readily seen  from panel c of Figure \ref{fig:temperature_vs_radii} where all three layers converge to a similar temperature, and in Figure \ref{fig:temperature_vs_vertical} where the outer vertical temperature profile is almost flat-topped.
This result is somewhat surprising in multiple ways. First of all, we did not observe this behavior in the tests performed on the synthetic disk models. In these tests, we varied the beam size and the disk inclination and yet always obtained a top and a bottom layer separated by a colder midplane (Appendix \ref{sec:model_testing}), indicating that this is not a feature of the TRD technique. Therefore there appears to be an additional process that leads to a much more vertically uniform temperature than expected. Second, a sharp jump in temperature is inconsistent with typical trends as a function of distance from the central star. This indicates that there is an important transition in the disk properties at this location. Finally, the 30\,K that is observed in the outer disk is much higher than the expected blackbody equilibrium temperature, which is about 20\,K and 15\,K at 200 and 300\,au, respectively, assuming low-albedo dust grains. Since this temperature has a very shallow dependency on stellar luminosity, it is highly unlikely that this could be due to the star being much more luminous than we assumed in our models. This points to the gas being overheated compared to expectations. Here we speculate on a possible explanation for these behaviors.

\begin{figure*}[ht!]
\plotone{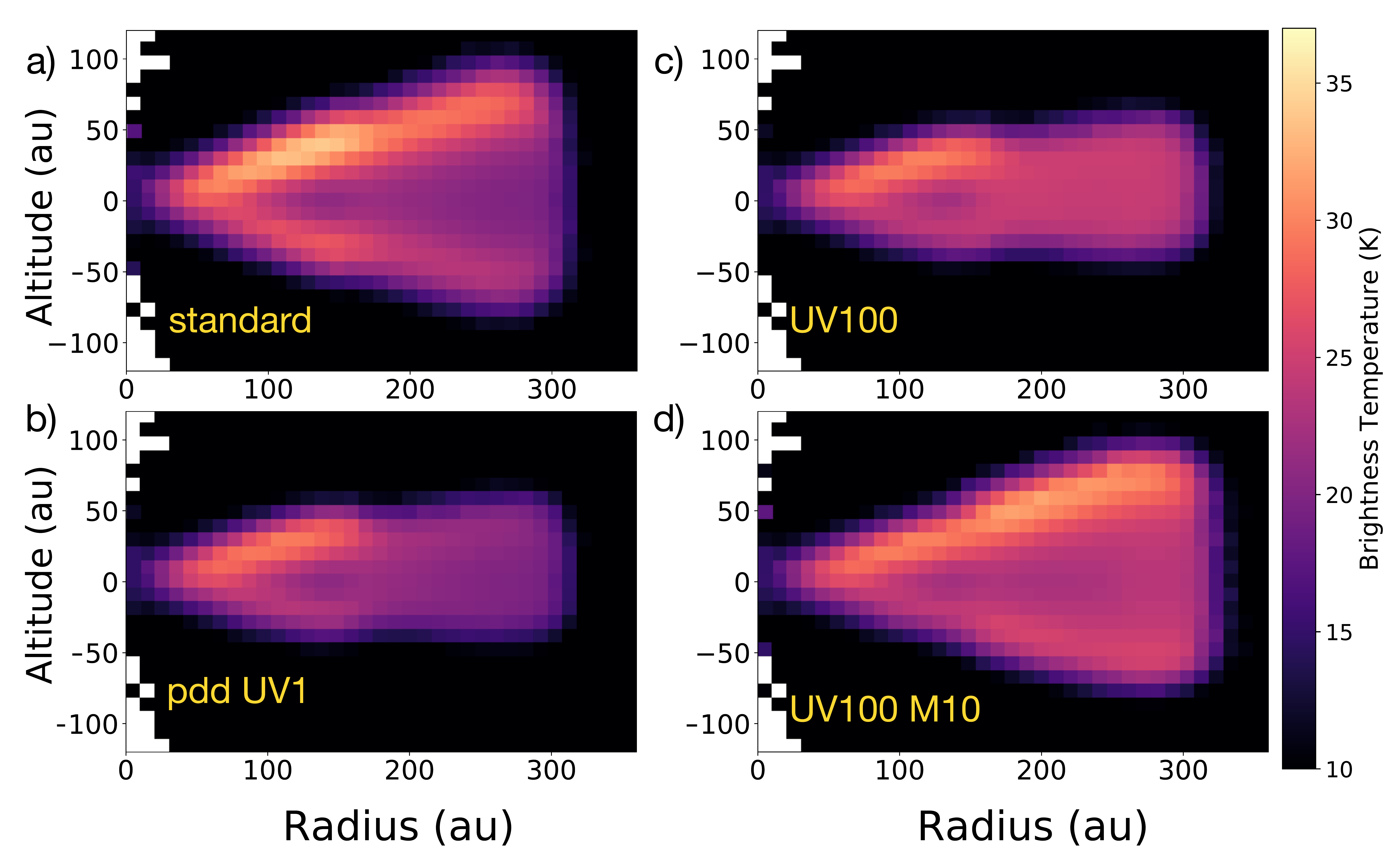}
\caption{MCFOST models with external UV irradiation might explain the flat-topped outer regions of Oph163131 as well as the increased temperature towards the outer parts of the disk. Panel a) shows the generic model without external irradiation. Panel b) shows the generic model but we turn on photodissociation and photodesorption with an UV irradiation environment equal to solar. Panel c) shows a 100 times increased UV irradiation environment. Panel d) shows the generic model with 100 times the solar UV irradiation (same as c), but the mass of the outer disk is increased by a factor of 10.}
\label{fig:increased_UV_models}
\end{figure*}

The radial position where the midplane temperature of Oph163131 increases from 21~K to 30~K, coincides with the position where the millimeter continuum emission is no longer detected (R$\sim200$~au, see Figure \ref{fig:HST_ALMA_cont} and Paper\,I). A similar behavior is seen in the Flying Saucer \cite{Dutrey2017}, in which a sharp (albeit smaller) temperature jump is observed at the outer radius of the millimeter continuum map.
However, one significant difference between the two disks is that submicron dust is traced in scattered light out to much larger radii in the Flying Saucer, whereas in Oph163131 the radial extent of the millimeter continuum and scattered light images are very similar to one another (see Paper\,I). We speculate that the dust grain difference relates to the observed midplane temperature profile outside of the millimeter continuum: it is uniform for Oph163131 but gradually declining for the Flying Saucer. 

\cite{Dutrey2017} proposed that this sharp temperature increase is related to a higher influx of UV radiation penetrating the disk from the outside, leading to enhanced thermal or direct CO desorption. This is supported by our observations of Oph163131.
Broadly speaking, the depletion of dust in the outer disk leads to a lower optical depth and thus higher penetration of short wavelength radiation. This process is even more dramatic in the case of Oph163131 where even small dust grains, which have high UV opacity, are absent or rare in the outer disk. In this case, we expect interstellar photons to penetrate deeper in the disk and to deposit more heat, leading to an even more uniform temperature in that region.

The much reduced dust density could also play a role in the heating of the outer parts as the gas could be more decoupled from the dust, allowing it to remain at a higher temperature. Large dust grains are known to be the most efficient emitters of far-IR and sub-millimeter radiation, and without them, the outer disk cannot radiatively cool as efficiently as in the inner parts. In this regard, we note that Oph163131 is positioned between the L1689 and L1688 Ophiuchus molecular clouds, in a region of substantially lower dust and gas column density than the Flying Saucer \citep{Lombardi2008, Ladjelate2020}. The remarkably uniform and warm temperature of the outer region of Oph163113 could thus be the consequence of enhanced interstellar UV radiation coupled with much lower dust opacity.

To go beyond pure speculations, we computed MCFOST synthetic disk models to test this hypothesis. We used the stellar parameter derived for Oph163131 and used disk geometrical and density parameters that visually matched the full extent of the disk around Oph163131. None of these models are meant to reproduce the observations of the gas component of Oph163131 but to demonstrate the effects of a higher external UV flux on the disks. We first constructed a `standard' model that consists of an inner region (R$<$170 au) with a dust mass of $10^{-3}$~$\rm M_\odot$ and an outer region with a dust mass of $10^{-6}$~$\rm M_\odot$ that extends from 170 au to 300 au. Both regions have a gas-to-dust ratio of 100, but the outer region contains only small grains (the maximum grain size was set to 10 \micron). This last requirement comes from the lack of sub-mm dust emission detected for Oph163131 at distances $>$200 au \citep[see Paper\,I and][]{Villenave2020}.  
Finally to perform a meaningful comparison to our results, we applied the TRD technique on this synthetic data, just as we did in Section \ref{susbsec:the_inner_disk} 

Figure  \ref{fig:increased_UV_models} shows the TRD maps for above mentioned models. The first model (panel a) corresponds to the `standard model', in which we did not include photodesorption or photodissociation of the CO molecule, as a purely illustrative exercise. This model shows two well separated top and bottom layers extending from the inner part of the disk and all the way to the outer disk. In this model, the temperature of the midplane in the outer parts is of the order of 17~K, while the top layer reaches a maximum temperature of 26~K, indicating a strong vertical gradient that is not seen in the data.  In the second model (panel b), we show the same `standard' model but now we have activated photodesorption and photodissociation in the model, and external illumination is implemented with a UV flux set to the UV field in the Solar neighborhood value \citep{Draine1996}. We see from this model that the interstellar photons have eroded the upper and bottom layers of the disk, creating a vertically flatter outer disk. Although the temperature of the midplane in the outer regions is still of the order of 17~K, we  notice that the temperature has an almost uniform value at all vertical elevations. The third model (panel c) is a variant of the same model where the UV irradiation field is 100 times stronger than the value in the Solar neighborhood. The increased irradiation field does not alter the shape of the disk any further but it homogeneously increases the temperature of the outer disk to 22~K. In this model, we see the distinctive feature of an outer disk's midplane that is hotter than the inner disk's midplane, i.e., a radial temperature increase in the midplane's temperature at radii larger than $\sim100$~au. Finally, in a fourth model (panel d), we used the same parameters as in the previous model, i.e., a UV flux 100 times the Solar neighborhood value, but we also increased the mass of the outer disk by a factor of 10, i.e., to $10^{-5}$~$\rm M_\odot$ in dust ($10^{-3}$~$\rm M_\odot$ in gas). As can be seen in Figure  \ref{fig:increased_UV_models}, the increased outer disk mass leads to self-shielding of the CO gas in the outer disk,
producing the formation of distinct upper and lower layers, a colder midplane and an overall flared appearance of the outer disk. The temperature of the top layer increased to 28~K, but the midplane temperature decreased to 20~K. Although it is not shown in the figure, we tested models with an even larger UV irradiation field. By the time the outer disk reached a temperature of 30~K, the radiation in the models was so high (1000 times the Solar neighborhood value) that it causes the whole disk to increase in temperature.

In conclusion, we showed that photodissociation and photodesorption are processes that can explain the vertically flat outer part of a disk, due to the erosion of the low-density top and bottom layers, as long as the overall gas column density (equivalently, mass) in the outer layer remains quite low, requiring a sharp decrease in surface density at 170--200\,au. The stripping of these layers produces in turn a vertically isothermal outer disk. In addition, when the UV irradiation field is increased (compared to Solar neighborhood values) it causes a rise in the temperature of the outer parts, allowing the possibility of having an outer disk midplane that is hotter than the inner disk midplane. 
Finally, we point out that to obtain optimal disk parameters for this source, it would be necessary to perform a detailed radiative transfer modeling of the CO, varying density and geometrical parameters, and including external UV irradiation along with photodesorption and photodissociation processes.

\section{Conclusion}
\label{sec:conclusion}
In this study, we present the first ALMA CO observations of the disk around Oph163131, as well as, the first optical and NIR spectroscopic observations of the central star. A summary
of our results are as follows:

\begin{itemize}
\item We obtained optical and NIR low-resolution spectra (R$\sim1000-2000$) of Oph163131 with the Goodman and SpeX spectrographs, respectively. By comparing the strengths of several absorption lines in both wavelength regions against appropriate stellar templates, we derived a spectral type of K4$\pm 1$\,IV for the star.
\item From the optical spectrum of Oph163131, we measured a $0.6\pm0.1$ \AA\, equivalent width of the H$\alpha$ line. This measurement in addition to a null detection of the Br$\gamma$ or Pa$\beta$ emission lines indicates that  Oph163131 is currently not accreting at an appreciable level. A detection of the He I line at 1.08 $\micron$, on the other hand, suggests that there might be either outflow activity or a very low accretion.
\item We used the velocity pattern imprinted in the disk by the star to infer a dynamical mass for Oph163131. Assuming a distance of 147 pc and an inclination angle of $i = 87 \degr$, we derived a stellar mass of $1.2 \, \rm M_{\odot}$ for the central source with conservative upper and lower mass limits of $1.4  \, \rm M_{\odot}$ and $1.0  \, 
\rm M_{\odot}$, respectively.
\item We have generalized the TRD method developed by \cite{Dutrey2017} to include non-perfectly edge-on disks. This generalization increases the number of disks that can be analyzed using this technique. In the Appendix, we have characterized some of the limitations of this method that can produce, for example, an apparent temperature decrease in the low density regions of the disk and a temperature increase of the top side of a disk (closest to the observer) in the presence of small inclinations.
\item Using the tomographically reconstruction distribution method, we obtained a model-independent measurement of the temperature structure of the disk surrounding Oph163131. Towards the inner R$<$200 au regions of the disk, we detect two emission layers separated by a dark lane, that we interpret as the top and bottom parts of a highly inclined disk separated by a cold and probably frozen-out midplane.
\item The colder midplane temperature of 21 K in the disk around Oph163131 increases to $\sim 30$~K for distances R$> 200$~au. This temperature increment coincides with a decrease in the micron and millimeter size dust detection in the disk. MCFOST models with increased UV irradiation qualitatively match the temperature increase, as well as, the vertically thinner structure of the outer regions. We propose that external UV irradiation could heat up the gas component in the outer disk, as the interstellar photons can more easily penetrate regions where dust particles are not present.
\end{itemize}

Overall, the disk surrounding Oph163131 is markedly more compact along the vertical direction than other well-studied disks, in both its dust and gaseous components. 
Combined with strong evidence for dust settling, low- to absent accretion on the central star, and low-density, over-heated outer region, this suggests that the system is a relatively advanced state of evolution that warrant further analysis. In particular, detailed line radiative transfer modeling of the CO maps presented here and new observations focused on different molecular tracers could provide a thorough picture of the vertical temperature profile, from the midplane to the upper surface of the disk.

\newpage

\acknowledgments
We thank the referee for an insightful review which helped to improve this paper. CF and MSC acknowledge support from the NASA Infrared Telescope Facility, which is operated by the University of Hawaii under contract 80HQTR19D0030 with the National Aeronautics and Space Administration, CF also acknowledge support from the Graduate Student Organization at the University of Hawaii at Manoa. 
GD acknowledges support from NASA grants NNX15AC89G and NNX15AD95G/NExSS as well as 80NSSC18K0442. 
KRS and DLP acknowledge support from HST GO grant 12514 from the Space Telescope Science Institute. FMe, MV, GvdP acknowledge funding from ANR of France under contract number ANR-16-CE31-0013. CF is grateful to Nienke van der Marel for an early mentoring that later developed into this publication.

\vspace{15mm}
\vspace{15mm}

%

\vspace{5mm}
\facilities{ALMA, JCMT, SOAR, IRTF}


\software{Astropy \citep{Astropy2018}, 
          CASA \citep{McMullin2007},
          MCFOST \citep{Pinte2006,Pinte2009},  
          Spextools \citep{Cushing2004}, 
          xtellcor \citep{Vacca2003}
          }


\bibliography{sample63}{}
\bibliographystyle{aasjournal}




\newpage

\appendix

\section{Equations to perform the Temperature Reconstruction of a non-perfectly edge-on disk}
\label{sec:Equations}
In this section, we provide the equations that relate a protoplanetary disk's velocity and geometrical coordinates as seen from an observer's point of view to the velocity and geometrical coordinates in the disk's coordinate system.

We define the disk's spatial coordinates 
as $\{ x,y,z \}$ and its velocity coordinates as $\{ v_x,v_y,v_z \}$. An observer at a finite distance $d$ and whose line of sight is inclined at an angle $\alpha$ with respect to the disk's midplane defines its own coordinate system $\{ x^\prime,y^\prime,z^\prime \}$ and velocity coordinates $\{ v_x^\prime,v_y^\prime,v_z^\prime \}$. Both coordinate systems are related by the following matrix transformation.

\begin{equation}
\label{eq:matrix_transformation}
\begin{pmatrix} 
x^\prime \\
y^\prime \\
z^\prime
\end{pmatrix}
=
\begin{pmatrix} 
1 & 0 & 0 \\
0 & \cos{\alpha} & -\sin{\alpha} \\ 
0 & \sin{\alpha} & \cos{\alpha} \\ 
\end{pmatrix}
\cdot
\begin{pmatrix} 
x \\
y \\
z
\end{pmatrix}
\end{equation}

We further assume that the disk is rotating at Keplerian speed $v=\sqrt{GM/r}$ and that the velocity component in the $y$ direction in the disk's reference frame can be described as:

\begin{equation}
\label{eq:disk_velocity}
    v_y= x\sqrt{\frac{\rm GM}{(\rho^2+z^2)^{3/2}}} \quad ; \quad \rho^2 = x^2 + y^2
\end{equation}

in the second part of equation (\ref{eq:disk_velocity}), $\rho$ represents the disk's radial distance in cylindrical coordinates as viewed from the disk's reference frame. We now combine equations (\ref{eq:matrix_transformation}) and (\ref{eq:disk_velocity}) to derive the line of sight velocity of the disk from the observer's perspective

\begin{eqnarray}
    v_y^\prime &=& v_y \cos{\alpha} - v_z \sin{\alpha} \quad ; v_z = 0 \\
    v_y^\prime &=& x \sqrt{\frac{\rm GM}{(\rho^2+z^2)^{3/2}}} \cos{\alpha} \label{eq:line_of_sight_velo_prime}
\end{eqnarray}

in the previous equation we assumed that the disk is not moving in the z-direction. We now would like to express the radial distance $\rho$ in the disk's rest frame as a function of the observer's coordinates $v_y^\prime, x^\prime,$ and $z^\prime$. From equations (\ref{eq:line_of_sight_velo_prime}) and (\ref{eq:matrix_transformation}) we have

\begin{eqnarray} 
\rho^2 &=& \left( \frac{x^{\prime 2}\rm{GM} \cos^2{\alpha}}{v_y^{\prime 2}} \right)^{2/3} -z^2 \\
\rho^2 &=& \left(\frac{x^{\prime 2}\rm{GM} \cos^2{\alpha}}{v_y^{\prime 2}} \right)^{2/3} - (z^{\prime}\cos{\alpha} - y^{\prime} \sin{\alpha})^2
\end{eqnarray}

which after some algebra we can write as:

\begin{eqnarray}
\label{eq:rho_equation}
\rho^2 = x^{\prime 2} + \left( z^{\prime} \sin{\alpha} \pm \cos{\alpha} \sqrt{gm - (x^{\prime 2} + z^{\prime 2})} \right)^2
\end{eqnarray}

with

\begin{equation}
gm = \left( \frac{x^{\prime 2}\rm{GM} \cos^2{\alpha}}{v_y^{\prime 2}} \right)^{2/3}
\end{equation}
\\
we take the real and positive solution of $\rho$, appropriate for the radius in cylindrical coordinate.

We additionally solve for the vertical geometrical component $z$ from the disk's coordinate system as a function of the observer's coordinates. Using equation (\ref{eq:matrix_transformation}), we have

\begin{eqnarray}
z &=& z^{\prime} \cos{\alpha} - y^{\prime}\sin{\alpha}\\
y^{\prime} &=& \frac{\sqrt{\rho^2 - x^{\prime 2}} - z^{\prime}\sin{\alpha}}{\cos{\alpha}}
\end{eqnarray}

which we combine to obtain

\begin{eqnarray}
\label{eq:z_equation}
z &=& z^{\prime} \cos{\alpha} - \left( \sqrt{\rho^2 - x^{\prime 2}} - z^{\prime} \sin{\alpha}\right) \tan{\alpha}
\end{eqnarray}

Therefore, for a given point in the PV diagram ($x^{\prime}$,$V_y^{\prime}$ space) and for a given $\alpha$ angle and $z^{\prime}$ vertical position, we can use equations (\ref{eq:rho_equation}) and (\ref{eq:z_equation}) to find the radial and vertical position of an emitting parcel of gas in the disk.

\section{TRD model testing}
\label{sec:model_testing}
In order to understand the limitations of the TRD method, we performed several tests on synthetic disk models. MCFOST is a line and continuum radiative transfer code that can produce atomic and molecular line maps of complex structures such as protoplanetary disks. To create a synthetic disk model, MCFOST first computes the temperature structure of a model assuming one or several energy sources. Under the assumption that dust and gas are thermally coupled (LTE), MCFOST then calculates the intensity of a given atomic or molecular species, creating a line map that can be compared to observations. In our models, we assumed that the dominant  source of energy is a single star with stellar parameters consistent with our results in Sections \ref{subsec:sepctral_characterization} and \ref{sec:dynamical_mass}. We assumed a tapered-edge disk geometry following \cite{Panic2009} and \cite{Pinte2018} with density and geometrical parameters qualitatively consistent with the observations. We emphasize that none of the disk models presented in this section are meant to be good a fit to the ALMA observations but to provide an idea of how different parameters such as the beam size of the observations and the inclination angle of the disk affect the TRD method. In these test models, we do not include photodesorption nor photodissociation effects. In Table \ref{table:model_disk_parameters}, we summarize the stellar and disk parameters we used to create the synthetic disk models.

\subsection{The ideal model}

\begin{figure}[ht!]
\epsscale{1.1}
\plotone{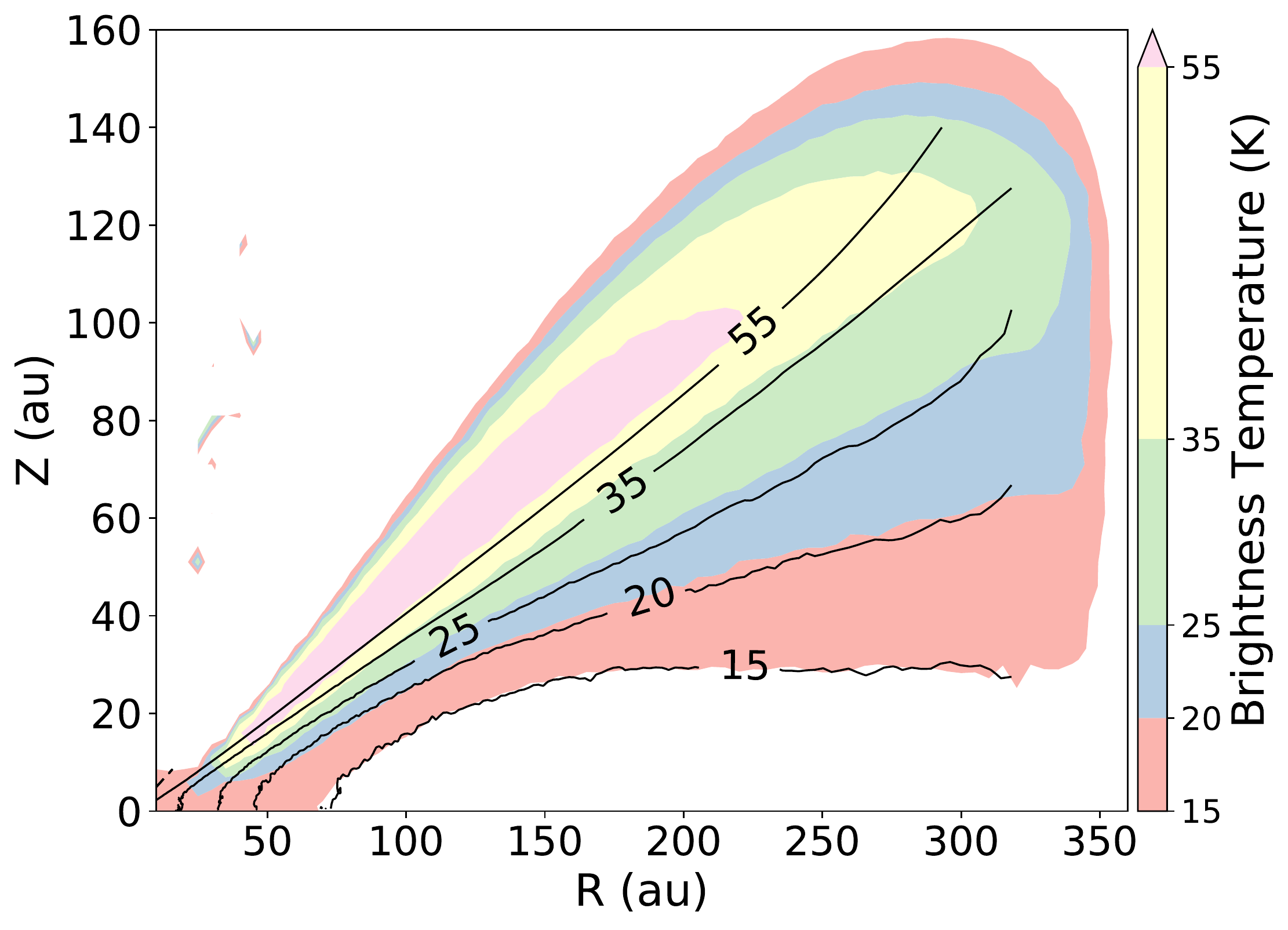}
\caption{Comparison between the TRD method applied to the synthetic disk model and the input disk temperature parameters. The model (shown in colors) was created with the parameters defined in Table \ref{table:model_disk_parameters} and convolved with a 0\farcs01 beam. The MCFOST input temperature shown as black lines range from 15~K to 55~K.}
\label{fig:TRD_method_perfect_model}
\end{figure}

The first synthetic model we created is a perfectly edge-on disk ($i$=90\degr) with an extremely high spatial resolution (0\farcs01 or 1.4~au) and a spectral resolution that matches our observations ($<0.25 \rm km s^{-1}$). Although we tried models with higher spectral resolution (0.01 $\rm km s^{-1}$), we found no significant differences between these two. We applied the TRD technique to this synthetic disk and compared the reconstructed temperature map to the input temperature from MCFOST in Figure \ref{fig:TRD_method_perfect_model}.
The input MCFOST temperature and the reconstructed temperature maps are shown as black lines and color maps, respectively with uneven steps of 5, 10, and 20~K to better display both temperatures. From the figure, we see that the temperature input and the retrieved temperature follow each other very closely at almost all radii and specially towards low-elevations. The 15~K to 35~K temperature regions, for example, match at almost all radii and elevations. The hotter temperature of 55~K, on the other hand, start deviating from the retrieved temperatures for radii larger than 200~au. The difference between the computed and the input temperatures is due to a lower density (lower optical depth ultimately) of the gas, which happens at both, very high elevations and larges distances from the center of the disk. In both cases, the reconstructed temperature is lower than the actual temperature of the disk. This behavior is expected and it follows from the definition of brightness temperature, which is known to coincide with the kinetic temperature of the gas when the optical depth of a line is sufficiently large. However, it is important to note that the models presented here do not have any included noise, therefore the optically thinner gas, often missed in the observations due to S/N detection limits, might not present a problem when using the TRD technique.

\begin{deluxetable}{lc}
\tablecaption{Disk and stellar parameters of the test model. \label{table:model_disk_parameters}}
\tablehead{
\colhead{Model Parameter} & \colhead{Value}}
\startdata
$a_{max}$ \, (\micron) &  1000 \\
$\rm M_{dust} \, (M_{\odot}) $& 5.5 10$^{-4}$  \\
gas/dust & 100 \\
CO/H$_2$ & $10^{-4}$ \\
$\beta$ & 1.25 \\
$\gamma$ & -1.0 \\
$R_{c}$ (au) & 40 \\
$h$ (au) & 15 \\
\hline
M ($\rm M_{\odot}$) & 1.2 \\
L ($\rm L_{\odot}$) & 0.96 \\
$\rm T_{eff}$ (K) & 4500 \\
\enddata
\tablecomments{The scale height $h$ is measured at 100 au.}
\end{deluxetable}

\subsection{TRD with different disk inclinations}
\label{subsec:models_different_inclinations}
The probability of encountering a protoplanetary disk that is perfectly edge-on is very small. It is therefore important to consider the cases where there are small deviations in the inclination of the disk towards the observers. 
To explore what are the effects of inclining the disk's out of edge-on position, we have computed two new MCFOST models with the same disk and stellar parameters as defined in Table \ref{table:model_disk_parameters} but this time we adopted inclinations of $i$=85\degr and $i$=80\degr.

\begin{figure*}[ht]
\plotone{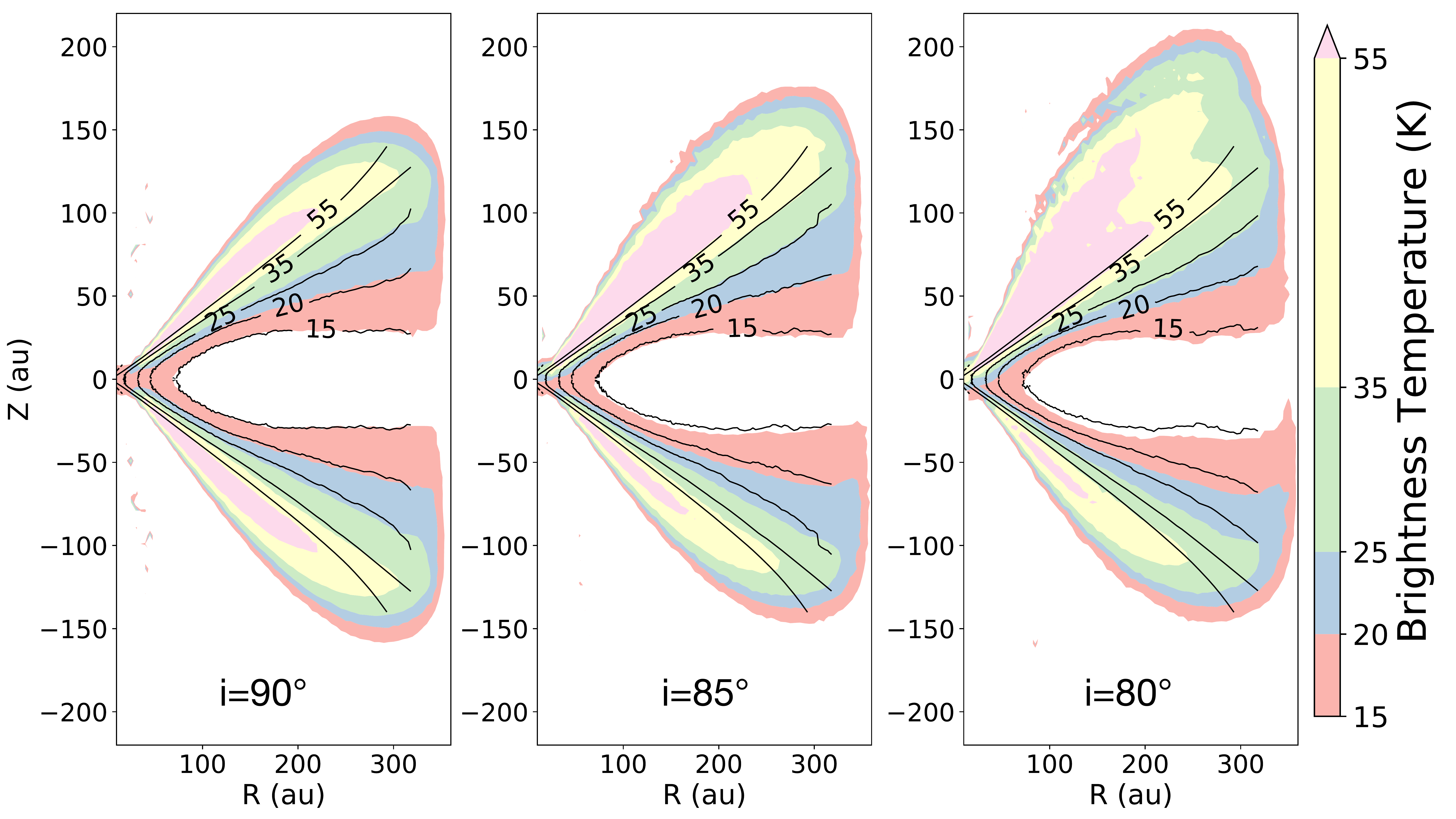}
\caption{Demonstration of the disk inclination variation when using the TRD method. From left to right we plotted the synthetic disk at inclination angles of $i=90\degr$, $i=85\degr$, and $i=80\degr$. The disks apparent temperature change with changing inclination in both, the layers above and below the midplane. Black contours lines correspond to the input MCFOST temperature.}
\label{fig:Changes_in_inclination}
\end{figure*}

In Figure \ref{fig:Changes_in_inclination}, we plotted these two new  models with inclinations of $i$=85\degr (middle) and $i$=80\degr (right panel), in addition to the default model with $i$=90\degr (left panel). This time both sides of the disk were plotted so the inclination effects can be visualized for the regions above and below the midplane. For the regions above the midplane, we see that as the disk is declined out of edge-on position, the far side of the disk becomes visible and lies above the photosphere of the disk overlapping with the near side. When the disk is inclined at $i$=85\degr, it is not possible to easily distinguish between the far and the near side of the disk in the reconstructed temperature map, so the emission from the far side is mixed with the emission from the photosphere on the near side, resulting in an increase on the measured temperature at higher elevations and also an increase in the scale height at all radii. When the disk is further declined to $i$=80\degr, the near and far side of the disk are clearly separated and the measured peak temperature returns to its original value. The layers below the midplane are also affected by changes in inclination. As the disk goes from $i=90\degr$ to $i=80\degr$ the near side gets hidden behind the disk's midplane causing that the light emitted from the bottom layer arises from a place closer to the midplane (because of optical depth effects) and therefore we see a cooler temperature.  

Although the TRD technique takes into consideration the inclination angle of the disk, the temperature reconstruction is affected by optical depth. This means, for example, that for a disk that is inclined at $i=80\degr$, the layer further from the observer (bottom layer in our Figure \ref{fig:Changes_in_inclination}) would not appear to be as hot as the closer layer (top layer), and this is not because the reconstruction is imperfect, instead this relates to the position in the disk where the optical depth of the line becomes much greater than unity. The same effect also causes that one side of the disk (top and bottom layers) appears to be larger than the other one. Although it might seem like that from Figure \ref{fig:Changes_in_inclination}, the reality is that we compare different temperature contours, which are arbitrarily selected. If one would allow the temperatures to range from much lower to much higher values, then the disk's would have the same sizes, but there would be temperature differences between them.


\subsection{TRD with different spatial resolution}
\label{subsec:models_different_resolutions}

Now we turn to the problem of how the spatial resolution affects the reconstructed temperature distribution of a protoplanetary disk. We use the same disk parameter as in the previous section (see Table\,\ref{table:model_disk_parameters}) and keep the inclination to $i$=90\degr. To simulate observations at different spatial resolutions, we convolved the synthetic disk model with a two dimensional circular Gaussian profile with a FWHM that varies from 0\farcs01 to 0\farcs3. 

\begin{figure*}[ht!]
\plotone{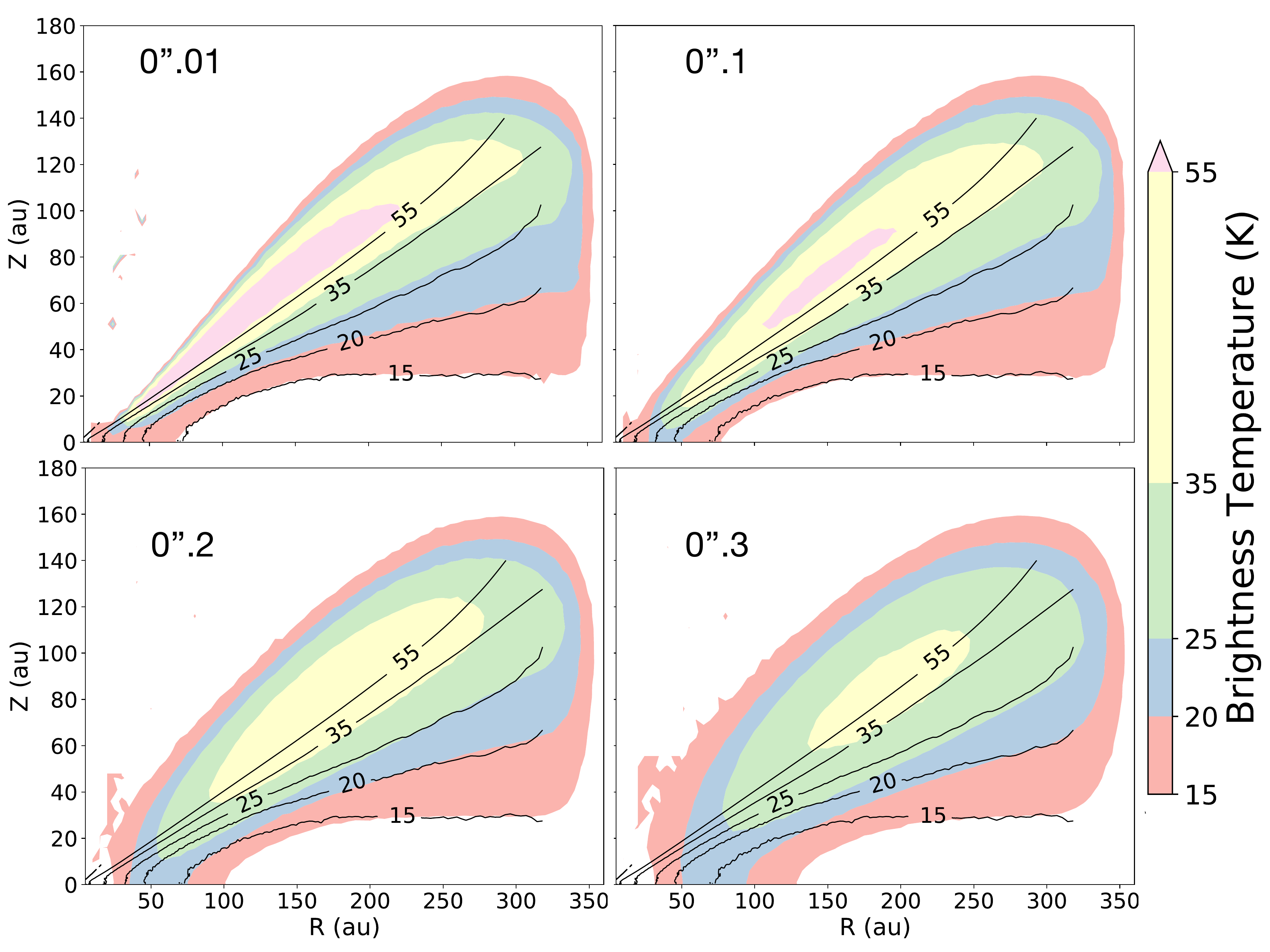}
\caption{Comparison between temperature retrieved model convolved with different beam sizes. The top left panel shows a model convolved with a 0\farcs01 beam (same as Figure \ref{fig:TRD_method_perfect_model}). Top right, bottom left, and bottom right panels show the same model convolved with a 0\farcs1, 0\farcs2, and 0\farcs3 beam sizes, respectively. Contours correspond to the input MCFOST temperature.}
\label{fig:Changes_in_spatial_resolution}
\end{figure*}

The spatial resolution effect is the most important limitation in our study of the temperature structure of a disk. It smears out the flux of the disk, limiting our ability to detecting high temperature regions confined in small scale structures. In Figure \ref{fig:Changes_in_spatial_resolution}, we illustrate this effect by convolving the same model shown in Figure \ref{fig:TRD_method_perfect_model} with different synthetic beam sizes. As we convolve the models with larger beam sizes, the retrieved peak temperature decreases, and the small scale emission layers disappear. When the spatial resolution of the observations is degraded, the retrieved peak temperature shifts towards the center of the disk. Additionally, the disk appears to be thicker in the vertical direction extending beyond the nominal disk's photosphere and also it becomes thicker towards the bottom part of the disk, where most of the CO has frozen-out. When the spatial resolution of the observations (for a disk of this size) is better than 0\farcs1, the convolution effects are much milder and the  retrieved temperature is comparable to the true temperature in the disk. It must be kept in mind, however, that even for very high spatial resolution observations, there are practical limits on the resolution of the reconstruction temperature maps. As the TRD method depends on the average of multiple pixels at a given radial position, on the shear Keplerian velocity, and on the turbulent velocity of the disk (as explained in Section \ref{sec:model})

To conclude, we would like to emphasize that if the spatial resolution of the observations is poor enough, the beam smearing effect largely exceeds the inclination effects. With the data we have presented, we have to acknowledge that inside of 125 au -- 150 au, we are likely to be affected by this effect, however, the outer disk is much less affected.

\end{document}